\documentclass[twocolumn]{aastex62}

\graphicspath{{../graphics/images/}{graphics/images/}{images/}}

\usepackage{subfiles}                   
\usepackage{amsmath}
\usepackage{amssymb}                    
\usepackage{booktabs}
\usepackage{color,soul}                 
\usepackage[caption=false]{subfig}
\usepackage{float}
\usepackage{color, colortbl}          
\usepackage{multirow}
\newcolumntype{Y}{>{\centering\arraybackslash}X} 
\renewcommand{\arraystretch}{1} 
\usepackage{balance}  

\definecolor{Gray}{gray}{0.9}

\DeclareMathOperator*{\argmax}{arg\,max} 

\shorttitle{Solar Image Parameter Data}
\shortauthors{Ahmadzadeh et al.}

\begin{document}

\title{A Curated Image Parameter Dataset from Solar Dynamics Observatory Mission}
\correspondingauthor{Azim Ahmadzadeh}
\email{aahmadzadeh1@cs.gsu.edu, dkempton1@cs.gsu.edu}

\author[0000-0002-1631-5336]{Azim Ahmadzadeh}
\affil{Georgia State University\\
Atlanta, GA 30302, USA}

\author[0000-0002-1837-8365]{Dustin~J.~Kempton}
\affiliation{Georgia State University\\
Atlanta, GA 30302, USA}

\author{Rafal~A.~Angryk}
\affiliation{Georgia State University\\
Atlanta, GA 30302, USA}

\begin{abstract}
We provide a large image parameter dataset extracted from the Solar Dynamics Observatory (SDO) mission's AIA instrument, for the period of January 2011 through the current date, with the cadence of six minutes, for nine wavelength channels. The volume of the dataset for each year is just short of 1 TiB. Towards achieving better results in the region classification of active regions and coronal holes, we improve upon the performance of a set of ten image parameters, through an in depth evaluation of various assumptions that are necessary for calculation of these image parameters. Then, where possible, a method for finding an appropriate settings for the parameter calculations was devised, as well as a validation task to show our improved results. In addition, we include comparisons of JP2 and FITS image formats using supervised classification models, by tuning the parameters specific to the format of the images from which they are extracted, and specific to each wavelength. The results of these comparisons show that utilizing JP2 images, which are significantly smaller files, is not detrimental to the region classification task that these parameters were originally intended for. Finally, we compute the tuned parameters on the AIA images and provide a public API\footnote{See: \url{http://dmlab.cs.gsu.edu/dmlabapi/}} to access the dataset. This dataset can be used in a range of studies on AIA images, such as content-based image retrieval or tracking of solar events, where dimensionality reduction on the images is necessary for feasibility of the tasks.

\end{abstract}




\section{Introduction}\label{sec:introduction}

\begin{table}
    \centering
    \caption{\footnotesize{The ten image parameters computed on the AIA images used to produce the dataset.}}
    \begin{tabular}{r | l  l}
        & Image Parameter & Formula \\
        \toprule
        1 & Entropy & \( - \sum_{i=0}^{L}{p(i) \cdot \log_2(p(i))}\) \\ \midrule
        2 & Mean ($\mu$) &  \( \sum_{i=0}^{L}{h(i)\cdot i}\)\\ \midrule
        3 & Standard Deviation ($\sigma$) & \( \sqrt{\sum_{i=0}^{L}h(i) \cdot (i - \mu)} \) \\ \midrule
        4 & Fractal Dimension & \( - \lim\limits_{\varepsilon \to 0}{ \frac{\log(N)}{\log(\varepsilon)}} \)  \\ \midrule
        5 & Skewness ($\mu_{3}$) & \( {\frac{1}{\sigma^{3}}} {\sum_{i=0}^{L}{h(i)(i - \mu)^3}} \) \\ \midrule
        6 & Kurtosis ($\mu_{4}$) & \( {\frac{1}{\sigma^{4}}} {\sum_{i=0}^{L}{h(i)(i - \mu)^4}} \) \\ \midrule
        7 & Uniformity & \( \sum_{i=0}^{L}{p^2(i)}\) \\ \midrule
        8 & Relative Smoothness & \( 1 - {\frac{1}{1+ \sigma ^2}}\) \\ \midrule
        9 & Tamura Contrast &  \( \frac{\sigma ^2}{{\mu_4} ^ {0.25}} \) \\ \midrule
        10& Tamura Directionality & See Eq.~\ref{eq:tdir}\\ \midrule
        \bottomrule
    \end{tabular}
    \begin{tabular}{p{0.3cm}p{7.0cm}}
        \\ \midrule
        $L$:  & maximum intensity value (e.g. $255$),\\
        $i$:  & color intensity value (\(i \in [0, L ])\), \\
        $p$:  & probability (i.e., normalized histogram),\\
        $h$:  & histogram,\\
        $N$:  & number of counting boxes,\\
        $\varepsilon$: & side length of the counting box\\
        \midrule
    \end{tabular}
    \label{table:tenParams}
\end{table}

Near real-time monitoring and recording of the Sun's activities has opened new doors for solar physicists to better understand the physics of different solar events. This was made possible in February $2010$, when the Solar Dynamic Observatory (SDO) \citep{2012SoPh..275....3P} was launched as the first mission of NASA's Living With a Star (LWS) Program, which is a long term project dedicated to the study of the Sun and its impact on human life \citep{withbroe:lws}. The SDO mission is invaluable for monitoring of space weather and prediction of solar events which produce high energy particles and radiation. Such activieis can have significant impacts on space and air travel, power grids, GPS, and communications satellites \citep{nrc:spaceweather}. SDO started capturing and transmitting to earth, approximately $70,000$ high-resolution images of the Sun, per day, or about $0.55$ petabytes of data per year \citep{martens2012computer}. This volume of data will only increase in time and with future missions. It is simply infeasible to take full advantage of such a large collection of data by traditional, human-based analysis of the images. Making it possible for solar physicists to extract information and knowledge from such a large volume of data, brings new challenges to other domains such as database management, computer vision, machine learning, and many others.\par

One of the primary objectives for improving the usability of such a large dataset is to reduce the size of the L1.5 FITS data without a significant loss of the information contained within the data. This can be done by utilizing either data compression algorithms or feature extraction (i.e., summarization) techniques, or both. While the features can be extracted from the highest quality of available data (in our study for instance, from AIA images in FITS format that we will discuss thoroughly later), the images may only be needed in smaller sizes or in compressed formats such as JP2000 or JPG. Of course, different approaches must be tailored for different tasks for which the data is being prepared, but an appropriate data reduction is extremely beneficial regardless.\par

By significantly reducing the size of the dataset, many useful tasks are made possible that previously may have been too costly to compute, if at all. To name a few, this would pave the road for a more efficient search and retrieval of images, clustering of similar regions of images across a wider temporal window, classification of solar events based on their regional texture, tracking of different events in time, and even real-time prediction of solar phenomena, for which the total computation time must comply with the streaming rate of the SDO images. Such reduction in size not only allows faster operations but also keeps the focus on some key aspects of the data, called features. Reducing the raw data into some important features is crucial owing to the fact that image repositories inherit the `curse of dimensionality' as every pixel is represented in one dimension. These high dimensional spaces are problematic as they may yield misleading results in any analysis that requires statistical significance, and this expands to affect almost all machine learning techniques \citep{trunk1979problem, hinneburg2000nearest, verleysen2005curse}. The curse is attributed to the situation where the growth in dimensionality of the data space is so fast that the number of available data samples cannot properly fill up the high dimensional space, which renders machine learning models powerless. Another important outcome of reducing the data volume is that by providing a more manageable data repository that can be easily accessed and managed by anyone without needing large and expensive storage devices or being highly skilled in dealing with `big data', more researchers from different domains may be encouraged to run different experiments on this collection of data and possibly provide more insight about the data.\par

To be able to more efficiently and accurately extract a set of important features from SDO's image data, various means of data mining should be utilized. This study builds upon a stack of techniques to derive the important image parameters, for the entire collection that is continuously being updated, starting from $2011$. Preprocessing of the original (L1.5) AIA image data, integrating the data with the spatiotemporal information such as the detected bounding boxes of different solar events' instances and the time stamp of their occurrences, extracting the important characteristics of the images, and labeling the instances are some of the major steps we take to transform the original data to the data that can be fed into the machine learning models. We utilize supervised learning to tune the features to reach their highest performance in classifying two important solar events' instances, namely active regions and coronal holes. In addition, we provide a comparative analysis between the extracted features from different image formats, in terms of their quality in distinguishing different solar events. In addition to providing the dataset as our primary goal, we hope that our detailed discussion on these topics would be informative for scientists interested in SDO images, or extraction of image parameters in general.\par

\begin{figure}[htp]
    \includegraphics[width=1.0\columnwidth]{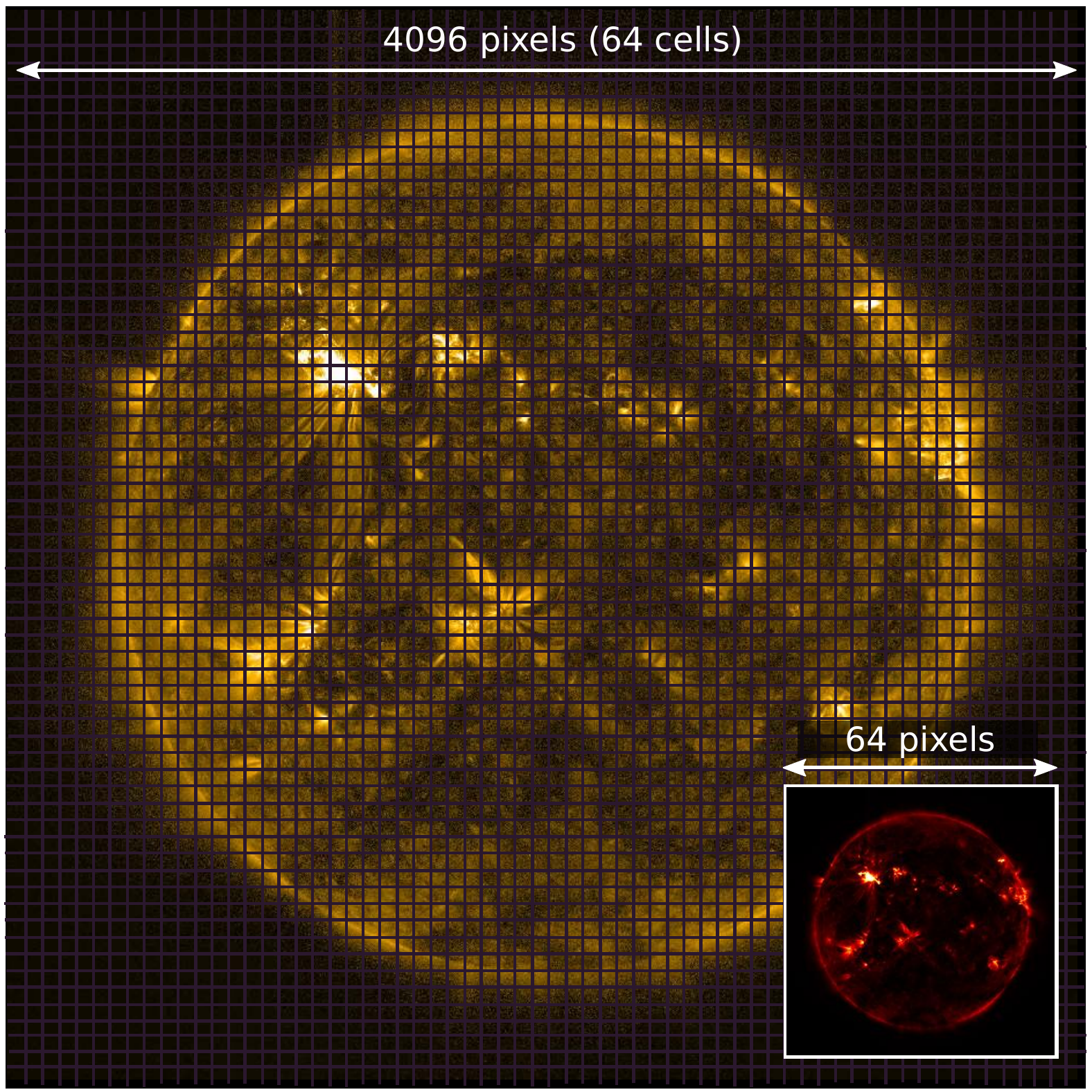}
    \caption{Grid-based segmentation of an AIA image with a grid of $64\times64$ cells, each of side length $64$ pixels. As an example, the mean image parameter is calculated on each cell and the resultant $64\times64$-pixel heat-map of the output is shown on the bottom-right corner. The heat-map is enlarged for a better visibility.}
    \label{fig:segmentation}
\end{figure}

Releasing the final dataset in the form of a public API will make the image-based analysis of the solar events easier and may open new doors to not only solar physicists but also computer scientists who are interested in feeding their models with a dataset different than the existing, general-purpose, image repositories.\par

\begin{figure*}[htp] 
    \subfloat[Entropy\label{fig:entropy}]{%
        \includegraphics[width=0.18\textwidth]{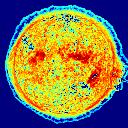}
    }\hfill
    \subfloat[Mean\label{fig:mean}]{%
        \includegraphics[width=0.18\textwidth]{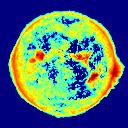}
    }\hfill
    \subfloat[Std. Deviation\label{fig:std}]{%
        \includegraphics[width=0.18\textwidth]{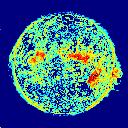}
    }\hfill
    \subfloat[Fractal Dim\label{fig:fdim}]{%
        \includegraphics[width=0.18\textwidth]{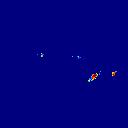}
    }\hfill
    \subfloat[Skewness\label{fig:skewness}]{%
        \includegraphics[width=0.18\textwidth]{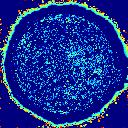}
    }
    \hfill
    \subfloat[Kurtosis\label{fig:kurtosis}]{%
        \includegraphics[width=0.18\textwidth]{graphics/images/Heatmaps/Skewness_AIA171.jpg}
    }\hfill
    \subfloat[Uniformity\label{fig:uniformity}]{%
        \includegraphics[width=0.18\textwidth]{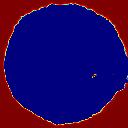}
    }\hfill
    \subfloat[Rel. Smoothness\label{fig:rsmoothnesss}]{%
        \includegraphics[width=0.18\textwidth]{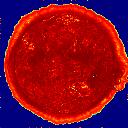}
    }\hfill
    \subfloat[T. Contrast\label{fig:tcontrast}]{%
        \includegraphics[width=0.18\textwidth]{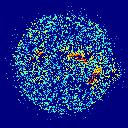}
    }\hfill
    \subfloat[T. Directionality\label{fig:tdir}]{%
        \includegraphics[width=0.18\textwidth]{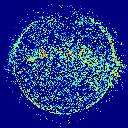}
    }\hfill
    \subfloat{%
        \includegraphics[width=1\textwidth]{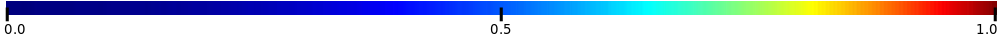}
    }
    \caption{Heatmap plots of the ten image parameters extracted from an AIA JP2 image captured on 2017-09-06 at 12:55:00, from the $171$--{\AA} channel.}
    \label{fig:heatmaps}
\end{figure*}

The remaining of this paper is organized in the following way: A background overview on SDO data and the image parameters that we are interested in is presented in Section~\ref{sec:background}. In Section~\ref{sec:datasources}, we explain the different sources we retrieve the data from and discuss the image types we run our models on. We then in Section~\ref{sec:settings}, analyze each of the image parameters and their variables which require tuning. The tuning process, and its evaluation using supervised learning, is presented in Section~\ref{sec:experimental}. After finding the best setting for each of the image parameters, we provide a thorough analysis of the produced data in Section~\ref{sec:thedata}. Section~\ref{sec:conclusion} concludes this work and discuss the future work. And finally, in \ref{app:one}, we present some statistical analysis of the created dataset to paint a more accurate picture of the reliability and usability of the data.\par

\section{Background}\label{sec:background}

The Solar Dynamic Observatory (SDO) was launched on February $11$, $2010$, as the first mission of NASA's Living With a Star (LWS) Program, with a five-year prime mission lifetime. The main goal of this project is to better understand the physics of solar variations that influence life and society. Now that it has been close to a decade since its launch, the observatory has provided us with approximately $4$ petabytes of data in total and is currently continuing to record even more. The Atmospheric Imaging Assembly (AIA), as one of the three SDO instruments, focuses on the evolution of the magnetic environment in the Sun's atmosphere and its interaction with embedded and surrounding plasma \citep{2012SoPh..275...17L}.\par

The AIA images archived in the Joint SDO Operations Center (JSOC) \footnote{JSOC; joint between Stanford and the Lockheed Martin Solar and Astrophysics Laboratory (LMSAL)} science-data processing (SDP) facility, have been processed by the SDO Feature Finding Team (FFT)\footnote{An international consortium groups selected by NASA to produce a comprehensive set of automated feature recognition modules.} \citep{martens2012computer} using its $16$ post-processing modules. The modules are designed for detection of solar event classes such as flares, active regions, filaments, and CMEs, in near real-time, and others such as coronal holes, sunspots, and jets. The results are posted at least twice a day to the Heliophysics Event Knowledgebase (HEK) system \citep{hurlburt2010heliophysics} since March 2010. One of the FFT's modules, which targets AR and CH events is called SPoCA suite \citep{verbeek:spoca}. SPoCA, or Spatial Possibilistic Clustering Algorithm, is run in near-real time at Lockheed Martin Solar and Astrophysics Laboratory and reports to the AR and CH catalogs of the HEK. It works on a variety of data sources including SDO's AIA images. SPoCA segments EUV images into three classes, namely, AR, CH, and QS. That is, it eventually attributes each pixel to one of the three classes, after running different fuzzy clustering algorithms on the images and applying some pre- and post-processing filters.\par

Due to the size of the dataset produced by the SDO, an efficient search and retrieval system over the entire archive is a necessity. In $2010$, this issue was first explored by Banda et al., and the ambitious task of creating a Content-Based Image Retrieval (CBIR) system on the SDO AIA images was started \citep{banda2010selection}. Given the volume and velocity of the data stream, the ten best image parameters (listed in Table~\ref{table:tenParams}) were chosen based on their effectiveness in classification of the solar events and also their processing time \citep{banda2010experimental}. The concern regarding the running time of the implemented parameters is rooted in the ultimate goal of near real-time processing of the data and the prediction of solar events. The processing window is therefore bounded by the rate of eight $4096\times4096$-pixel images being transmitted to earth every $10$ seconds. The performance of these parameters was further experimented and confirmed by \citet{banda2011surprisingly,banda2013steps}. Due to the variety of issues that must be addressed for a reliable CBIR system to be created, this is still an active research with the latest update in \citet{schuh2017region}.\par

In addition to the analysis performed in the previously mentioned works, these parameters have also been used for the classification of filaments in H-alpha images from the Big Bear Solar Observatory (BBSO) and similar success was reported by \citet{schuh2014massive}. Schuh et al. also employed these ten image parameters for the development of a trainable module for use in the CBIR system \citep{schuh2015solar}, along with a thorough analysis on three years of SDO data (from Jan $1$, $2012$ through Dec $31$, $2014$).  Yet another sequence of studies benefits from the same set of image parameters for tracking of the solar phenomena in time  \citep{kempton2015tracking, kempton2016towards, kempton:phenomena}. In that work, their tracking model utilize sparse coding to classify solar event detections as either the same detected event at a later time or an entirely different solar event of the same type. This model links the individually reported object detections into sets of object detection reports called tracks, using a multiple hypothesis tracking algorithm. This was accomplished through the consideration of the same set of image parameters on which we concentrate in this study. We hope that our thorough analysis, which results in a significant improvement in effectiveness of the ten image parameters, helps all of the above studies in their performance noticeably.\par

\subsection{Image Parameters}\label{subsec:imageparameters}

All parameters in Table~\ref{table:tenParams}, except for fractal dimension and Tamura directionality, capture some information about the distribution of the pixel intensity values of the images and none of them preserve the spatial information of the pixels. Even though the spatial information is not preserved, the distribution-related data provide many clues as to the characteristics of the image. For example, a narrowly distributed histogram indicates a low-contrast image. A bimodal distribution often suggests that the image contains an object or a region with a narrow amplitude range against a background of differing amplitude. However, the location and shape of the solar phenomena, similar to the temporal information, are the crucial aspects of our data. In order to help preserve some of the spatial information of the data, we apply a grid-based segmentation on the images. This is a widely used technique already experimented on the AIA images by \citet{banda2009effectiveness, banda2010experimental} that has shown good results. Each $4096\times4096$-pixel AIA image is segmented by a fixed $64\times64$-cell grid. For each grid cell that spans over a square of $64\times64$ pixels of the image, the $10$ image parameters will be calculated. In Fig.~\ref{fig:segmentation}, such segmentation, as well as the heat-map of the mean parameter ($\mu$) as an example, is visualized. Since we are processing $10$ parameters for each image, (see Fig.~\ref{fig:heatmaps}), the image then forms a data cube of size $64\times64\times10$. Additionally, for each time step, we process $9$ images from different wavelength filter channels of the AIA instrument.\par

The image parameters can also be categorized in two main groups; those which describe purely statistical characteristics of an image and those that capture the textural information. The former further divides into two subcategories: 1) Parameters such as mean, standard deviation, skewness, kurtosis, relative smoothness, and Tamura contrast that solely depend on the pixel intensity values of the image, 2) Parameters such as uniformity and entropy, that, in addition to the pixel values, depend on the choice of the bin size required for construction of the normalized histogram of the color intensities \footnote{Note that in Table~\ref{table:tenParams}, in order to have a unified formulation for different parameters, whenever possible we used the histogram function (i.e., $h(i)$) to formulate the parameter, however, it is only for two parameters, namely uniformity and entropy, that the calculation of the normalized histogram (i.e., $p(i)$) is necessary.}. The latter captures the characteristics of the image texture within the regions of interest (i.e., solar events). In the following text, we elaborate more on the four image parameters which require a deeper attention.\par

\subsubsection{Entropy}\label{subsubsec:entropy}

Entropy, as an image parameter, has been widely utilized in a variety of interdisciplinary studies ranging from medical images \citep{pluim2003mutual} to astronomical \citep{starck2001entropy} and satellite \citep{barbieri2011entropy} images. Depending on the specific goal in each study, different approaches might be needed. All of the suggested models try to measure the disorder or uncertainty of pixel values in an image (or bits of data in general). Almost all of them are inspired, one way or another, from the definition of entropy introduced by \citet{shannon2001mathematical} of the Information Theory domain. Despite the valuable achievements in this direction, the Monkey Model Entropy (MME) \citep{justice1986maximum,skilling1989classic} which is identical to what Shannon introduced for decoding communication bits, is still the most popular model in the image processing community. In this model, the random variable $i_{x,y}$, i.e., the intensity value of the pixel at position $(x,y)$, is assumed to be independent and identically distributed (i.i.d) and therefore the entropy is measured as follows:
\begin{equation}\label{eq:entropy}
\text{entropy}_{\text{MME}} = - \sum_{i=0}^{L}{p(i) \cdot \log_2(p(i))}
\end{equation}
where $p$ is the probability distribution function of the pixel intensity value $i$, and $L$ is the number of gray levels minus one (e.g., $255$ for a typical 8-bit quantized image). This can be computed directly from the intensity-based histogram of an image. As an intuitive interpretation of this parameter, one could say that an image with low entropy is more homogeneous than one with higher entropy.\par

This model of entropy was utilized previously by Banda et al., as one of ten selected image parameters in their research \citep{banda2010selection}. It is worth noting that we are aware of the fact that the assumption of i.i.d pixel intensities disregards the presence of spatial order or contextual dependency of the image pixels, however, the segmentation step discussed above provides some compensation for this loss of spatial information. In addition, the simplicity of this model is in line with the previously discussed focus on prioritizing the computation cost of the parameter choices. The MME is indeed the simplest model and can be computed faster than other approaches, for instance, those which require the computation of the joint probability distribution function of the pixel values \citep{razlighi2009comparison}.\par

\subsubsection{Uniformity}\label{subsubsec:uniformity}

Similar to entropy, uniformity is also a popular statistical measure that is widely used to quantify the randomness of the color intensities and to characterize the textural properties of an image. Uniformity is calculated as:
\begin{equation}\label{eq:uniformity}
    \text{uniformity} = \sum_{i=0}^{L}{p^2(i)}
\end{equation}
and reaches its highest value when gray level distribution has either a constant or a periodic form \citep{davis1979texture}. In this formula, the variables $p$, $i$, and $L$ are similar to those in Eq.~\ref{eq:entropy}, where $p$ is the probability distribution function of the pixel intensity value $i$, and $L$ is the number of gray levels minus one.\par

\subsubsection{Fractal Dimension}\label{subsubsec:fractalDimension}

Fractal dimension is another well-known measure utilized by scientists of different domains. However, unlike the parameters discussed so far which are purely statistical measures, fractal dimension (and Tamura directionality) focus more on the textural aspects that we believe are in particular importance for distinction of at least some of the solar phenomena, such as active regions and coronal holes. Whenever it comes to analyzing scientific image data, this parameter seems to be a useful choice. In solar physics, as a relevant example, fractal dimension was used for a variety of purposes including detection of active regions \citep{revathy2005fractal}, and to exhibit fractal scaling of solar flares in EUV wavelength channels \citep{aschwanden2008solar}.\par

Historically, fractal dimension was once used as a clever solution to a problem that is now known as the coastline paradox \citep{weisstein2008coastline}. It was the idea of measuring the length of the coast of Britain, independent from the scale of measurement \citep{mandelbrot1967long}, that provided the basis for the definition of this parameter. Fractal dimension is a measure of nonlinear growth, which reflects the degree of irregularity over multiple scales. In other words, it measures the complexity of fractal-like shapes or regions. A larger dimension indicates a more complex pattern while a smaller quantity suggests a smoother and less noisy structure. Among the several different methods for measuring the fractal dimension \citep{annadhason2012methods}, the box counting method, also known as Minkowski-Bouligand dimension, is the most popular one.\par

The general approach for the box counting method can be described as follows. The fractal surface, in an $n$-dimensional space, is first partitioned with a grid of $n$-cubes with the side length of $\varepsilon$. Then, $N(\varepsilon)$ is used to denote the number of $n$-cubes overlapping with the fractal structure. The counting process is then repeated for the $n$-cubes of different sizes, and the slope $\beta$ of the regression line fitting the plot of $\varepsilon$ against $N(\varepsilon)$ gives the dimension of this fractal. In a $2$-D space such as ours, the $n$-cubes are simply squares with a side length of $\varepsilon$. More details of employing this parameter for measuring the complexity of solar events is discussed in Section~\ref{sec:settings}.\par

\subsubsection{Tamura Directionality}\label{subsubsec:tamuraDirectionality}

Directionality as a texture parameter is a well-known concept in image processing and texture analysis domains. This parameter was extensively investigated by  \citet{bajcsy1973computer} and later on by \citet{tamura1978textural}. The proposed method by Tamura, used to measure the directionality, has become a popular texture parameter and has been used in a variety of studies. The well-known examples are in QBIC \citep{flickner1995query} and Photobook \citep{flickner1995query} projects which are content-based image retrieval (CBIR) systems. Some more domain specific examples would be the solar image data benchmark gathered by \citet{schuh2014massive} and the tracking of the solar events by \citet{kempton2015tracking}. In addition to Banda's work \citep{banda2010selection} on evaluating the effectiveness of Tamura directionality on AIA solar images, \citet{islam2008geometric}, a discipline-independent study, showed that directionality is indeed one of the most important texture features when the human perception is considered the ground truth.\par

Tamura directionality is a measurement of changes in directions visually perceivable in image textures. Tamura formulated this parameter as follows:

\begin{equation}\label{eq:tdir}
    T_{dir} = 1 - r \cdot n_{p} \cdot \sum_{p}^{n_{p}}{ \sum_{\phi \in \omega_{p}}^{}{(\phi - \phi_p)^2} \cdot h(\phi) }
\end{equation}

\noindent where: \newline
$p$: a peak's index,\\
$n_{p}$: the total number of peaks,\\
$\phi_{p}$: the angle corresponding to the $p$-th peak,\\
$\omega_{p}$: a neighborhood of angles around the $p$-th peak,\\
$r$: the normalizing factor for quantization level of $\phi$,\\
$\phi$: the quantized direction code (cyclically in modulo $180^\circ$).\\

In the statistical terms, this parameter calculates the weighted variance of the gradient angles, $\phi$, for each peak, $p$, of the histogram of angles, $h(\phi)$, within each peak's domain, $\omega_p$, considering the angle corresponding to each peak be the mean value of the angles within that peak's domain. It then aggregates across the identified peaks, and after re-scaling the result to the range $[0,1]$, it subtracts the final value from one to achieve a monotonically increasing function. That is, it returns greater quantities for a more directional texture.\par

\section{Data Sources}\label{sec:datasources}

In order to tune the calculation of image parameters for achieving an effective set of features requires an evaluation process. The evaluation process we utilize relies on reported solar events to evaluate the performance of each image parameter individually for each wavelength channel we are utilizing. In order to accomplish this, we use supervised learning to measure the performance of each of the image parameters in detecting some of the solar events. In this section, we detail our data sources for our images and the event-related metadata that was collected. We also briefly explain the FITS format, a commonly used format in astronomy that is employed by the SDO repository as the primary way for digitizing the AIA images. Understanding of the structure of this format and how the AIA images are stored in such format is crucial for our preprocessing steps.\par

\subsection{HEK: Event Data}\label{subsec:hekdata}

The Heliophysics Event Knowledgebase (HEK) is the source of the spatiotemporal data used in this study. The HEK system, as a centralized archive of solar event reports, is populated with the events detected by its Event Detection System (EDS) from SDO data. There are considered $18$ different classes of events such as active region, coronal hole, and flare. For each event class, a unique set of required and optional attributes is defined. Each event must have a duration and a bounding box that contains the event in space and time. We use this information to map the meta data of the reported events to the corresponding AIA images.\par

For the evaluation of image parameters performed in this study, we utilize two of the reported solar event types active region and coronal hole. There are multiple reporting sources for active regions that are reported to HEK, and those reported by the Space Weather Prediction Center (SWPC) of NOAA (National Oceanic and Atmospheric Administration) are assigned numbers daily. The NOAA active region observations, as \citet{hurlburt2010heliophysics} explains, is an event bounded within a $24$-hour time interval, and therefore HEK considers all NOAA active regions with the same active region number to be the same active region. However, there is a second automated module from the Feature Finding Team that reports both active region and coronal holes described by \citet{verbeek:spoca} and called the SPoCa module, which reports detections every four hours. It is the reports from this module that are utilized as the solar events of interest in this study.\par

\subsection{SDO: AIA Image Data}\label{subsec:sdodata}

The atmospheric imaging assembly (AIA) has four telescopes that provide narrow-band imaging of seven extreme ultraviolet (EUV) band passes ($94$--{\AA}, $131$--{\AA}, $171$--{\AA}, $193$--{\AA}, $211$--{\AA}, $304$--{\AA}, and $335$--{\AA}) and two UV channels ($1600$--{\AA} and $1700$--{\AA}) \citep{2012SoPh..275...17L}. The captured $4k$ images of the Sun, which are full-disk snapshots with the cadence of $12$ seconds, are compressed on board and without being recorded on orbit, are transmitted to SDO ground stations. The received raw data (Level 0) are archived on magnetic tapes in JSOC science-data processing facility. The uncompressed data is then exported as FITS files with the data represented as $32$-bit floating values. At this point, images are already calibrated, however, some corrections and cleaning are still required due to the existence of a small residual roll angle between the four AIA telescopes. At this stage (Level 1.5), the data is ready for analysis. In some repositories including Helioviewer, the FITS files are converted to JP2 format to reduce the volume of their database. In this study, we use the level $1.5$ (in short L1.5) FITS files and the JP2 images to achieve a comparative analysis. In the following subsections, we elaborate more on how FITS files are different from the JP2 images and why a fair comparison should take into account the differences in the distribution of pixel intensities in these two image formats.\par

\subsubsection{AIA Images in FITS}

\begin{figure*}[!t] 
    \includegraphics[width=\textwidth,height=2cm]{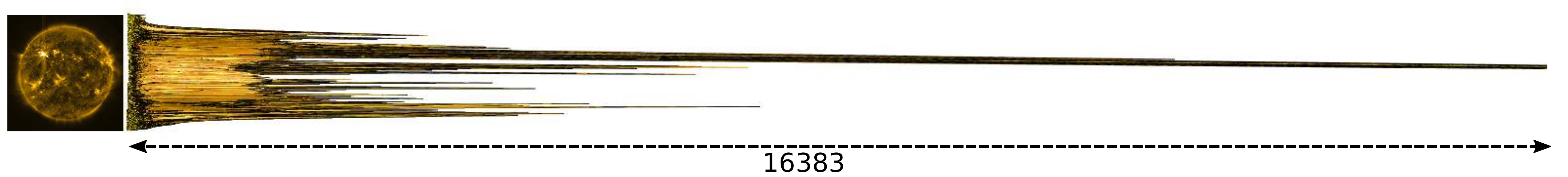}
    \caption{A 3-D view of an AIA FITS image from the $171$--{\AA} channel, with values ranging from $0$ to $16383$.}
    \label{fig:3d}
\end{figure*}

FITS, short for Flexible Image Transport System, is a data format for recording digital images of scientific observations. This format was proposed as a solution to the data transport problem. For details on FITS format we refer the interested reader to \citet{wells1979fits}. Here, we only mention a few key points about this format to provide the basic knowledge needed for understanding the preprocessing steps that will be discussed later. For processing of the FITS files we use the \textit{nom-tam-fits}\footnote{Library: \url{http://nom-tam-fits.github.io/nom-tam-fits/}} Java library.\par

A FITS file consists of a header where the basic and optional meta data are stored, and immediately following that is the data matrix representing the image starts. In the header of AIA images, a plethora of information is stored \citep{nightingale2011aia} that might be useful for different purposes, such as the minimum and maximum color intensities, the date of creation of the file, the exposure time of CCD detectors of the AIA instrument, the name of the telescope (e.g., SDO/AIA) and the instrument (e.g., AIA), wavelength in units of {\AA}ngstroms (e.g., $94$--{\AA}), several descriptive statistics about the captured intensities, radius of the Sun in pixels on the CCD detectors, and so on. It is important to note that unlike the typical 8-bit quantized image formats such as JP2, JPG, or PNG, that are limited to $256$ different intensity levels, the intensity level in FITS format is only bound to the sensitivity of the sensors of the camera. Since the AIA cameras use a $14$-bit analog-to-digital converter (ADC) to translate the charge read out from each pixel to a digital number value \citep{boerner2011initial}, the FITS color intensity value has an upper-bound at $16384$ (i.e., $2^{14}$). Such a level of precision comes at the cost of introducing a significant degree of skewness in the distribution of intensities. In the next section, this will be discussed in greater detail.\par

\subsubsection{Distribution of Pixel Intensities}\label{subsubsec:distribution}

Since in this study, we run all of our experiments on both JP2 and FITS images, it is important to have a good understanding of the distribution of pixel intensities in these two formats, the differences and similarities. We begin the discussion with the theoretical pixel intensity extrema in FITS files, i.e., $0$ and $16383$. For instance, in FITS format, the appearance of pixels with the maximum brightness is not as frequent as it is in the JP2 images.This is of course, the result of the JP2 lossy compression which transforms the pixel intensity domain of the FITS file into a much narrower range of $0$ to $255$. However, these extreme values are very likely to appear in FITS images, in the bright regions caused by the strong flares. In the other extreme, for FITS format images, some negative values might be present, which appear to be a byproduct of the post-processing data transformation (level $0$ to level $1.5$) since the CCD detectors are not capable of recording negative values. As a pre-processing step, we replace all the negative values with zeros in order to clean the data. It is interesting to note that such an extreme skewness in the distribution of pixel intensities is not limited to a specific wavelength channel, and is held true across all EUV and UV channels.\par

Next, we would like to learn about the amount of contribution of the extreme values in the distribution of pixel intensities. In this, we are interested in knowing the percentage of pixels in each image that carry such extreme values. To answer this question, we studied one month worth of AIA FITS images, since $2010.09.01$ through $2010.09.30$, with the cadence of $2$ hours, from $9$ wavelength channels (excluding the visible wavelength, $4500$--{\AA}), that sums up to a total of $3240$ images. In Fig.~\ref{fig:percentiles}, the $p$-th percentile of the observed intensities for each of the images within this period is shown. The maximum values in these plots should be compared against the maximum intensity reached during this period, which is the theoretical maximum, i.e., $16383$ for all $9$ wavelength channels. By looking at the spike in the first plot (i.e., wavelength $94$--{\AA}), we can see that $99.5\%$ of the pixels in the corresponding image had color intensities less than $44$, while pixels as bright as $16383$ existed in that very image. Such significant gaps between the mean values of the distributions and the maxima is summarized in Table~\ref{table:percentiles}.\par

\begin{figure}[htp] 
    \includegraphics[width=1.0\linewidth]{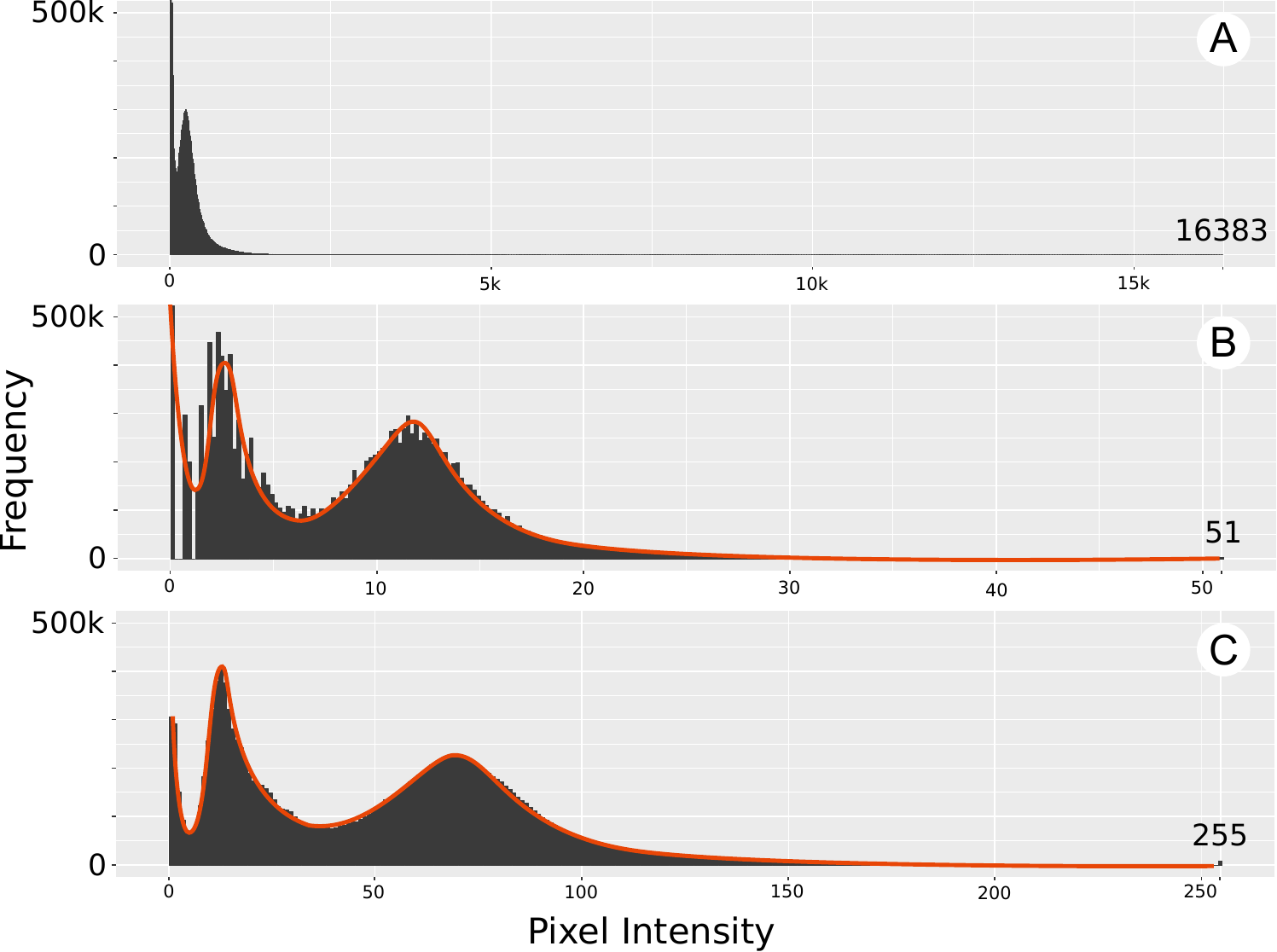}
    \caption{Distribution of pixel intensities in a FITS image (A), a clipped FITS (B), and in a similar image in JP2 format (C). The illustration shows how clipping of the raw FITS image can reveal the hidden shape of the bimodal distribution which is not visible in (A) due to the large number of bins. 
    }
    \label{fig:distributions}
\end{figure}

The above statistical analysis suggests an extreme skewness in the distribution of pixel intensities in FITS images. This is illustrated in plot A of Fig.~\ref{fig:distributions}. The visual effect of such skewness is ``underexposure''. In other words, if the pixel values of a FITS image are (linearly) transformed to the range of $8$-bit images (i.e., $[0,255]$), the output would be mostly black, with few to no small, extremely bright regions. It is important to note that our image parameters, which are utilized in supervised machine learning models to distinguish the different solar phenomena, are pixel-based features. That is, the relative differences between the pixels' brightness will be taken into account and not their absolute values. Therefore, providing the classifiers with the original L1.5 AIA data containing such far-out values, and not treating the outliers appropriately could bias the fit estimates and distort the classification results. We provide more details on how this issue is addressed in the next section.\par

\subsection{FITS, Clipped FITS, and JP2}\label{subsec:allFormats}

In this section, we will explain how we preprocess FITS files prior to the feature extraction and classification tasks. It is worth noting that, since such preprocessing steps introduce some changes on the pixel values of the original L1.5 FITS files, for the sake of completeness of our later comparisons and to avoid any bias in our study, we extend our experiments to cover the three data types: JP2, L1.5 FITS, and clipped FITS, as defined in the following sections.\par

\subsubsection{Clipping FITS Images}

Treating the outliers is a common practice in the process of cleaning the data for any machine learning task, as they may introduce a significant bias to the learning process and hence reduce the effectiveness of the extracted features for the classification goal. In the case of outliers being the extreme values, the general approaches are: a) removal of the outliers, b) replacing them with some statistics (imputation), c) altering with expected extrema (capping), and d) predicting their ``expected'' values based on the local changes of the intensities. Of course, the removal of the outliers and re-scaling the values into the quantized range of 8-bit values would leave us with the results not so much different than the existing JP2 images. This would void our attempt to study the potential differences in analysis of FITS versus JP2.\par

So, instead of removing outliers all together, we will employ the capping approach, that is also known as \textit{clipping} if applied to images. The process involves finding a global cutting point on the skewed tail of the probability distribution function, and shift all the pixel intensities above this threshold to this point. By ``global'' cutting points, we mean thresholds that are fixed across all AIA images for each wavelength channel. This ensures that the clipping filter affects all of the images uniformly. The result of such data transformation is that while no data points are removed (but shifted to the cutting point), the extreme skewness of the distribution is slightly mitigated. We use the maximum of the $99.5$-th percentiles of pixel intensities as the global cutting point for each wavelength. That is, in the worst case scenario, $0.5\%$ of the observed pixel intensities will be shifted to the new maximum point. The chosen cutting points for each wavelength is highlighted in Table.~\ref{table:percentiles}.\par

\newcolumntype{g}{>{\columncolor{Gray}}r}
\begin{table}
    \centering
    \caption{Maximum percentiles of the pixel intensities of AIA FITS images, observed from $9$ wavelength channels, for the period of $2010.09.01$ to $2010.09.30$, with the cadence of $2$ hours.}
    \begin{tabular}{r | r r r r g r}
        \textbf{W}  & \textbf{80-th} & \textbf{90-th} & \textbf{95-th} & \textbf{99-th} & \textbf{99.5-th} & \textbf{Max}\\ \toprule
        \textbf{94{\AA}}   & $7$       & $10$      & $15$      & $34$      & $44$      & $16383$\\ \midrule
        \textbf{131{\AA}}  & $19$      & $30$      & $43$      & $88$      & $123$     & $16383$\\ \midrule
        \textbf{171{\AA}}  & $568$     & $777$     & $1034$    & $1935$    & $2602$    & $16383$\\ \midrule
        \textbf{193{\AA}}  & $574$     & $904$     & $1354$    & $2884$    & $3968$    & $16383$\\ \midrule
        \textbf{211{\AA}}  & $154$     & $258$     & $429$     & $1159$    & $1673$    & $16383$\\ \midrule
        \textbf{304{\AA}}  & $116$     & $151$     & $188$     & $327$     & $431$     & $16383$\\ \midrule
        \textbf{335{\AA}}  & $16$      & $26$      & $43$      & $171$     & $305$     & $16383$\\ \midrule
        \textbf{1600{\AA}} & $196$     & $242$     & $289$     & $427$     & $509$     & $16046$\\ \midrule
        \textbf{1700{\AA}} & $1801$    & $2205$    & $2558$    & $3517$    & $4138$    & $16215$\\ \midrule
        \bottomrule
    \end{tabular}
    \label{table:percentiles}
\end{table}

\subsubsection{Pixel Intensity Transformation}\label{subsubsec:transformation}

After having used the statistically derived cut-off points for capping outlier pixel values, an additional processing step that should be done is to re-scale the now capped values. Note that after clipping the FITS images, although the distribution of pixel intensities are now more naturally skewed, they do not have the same distribution as the pixels in JP2 images have. This is due to the non-linear transformation of the data in conversion of FITS to JP2 format. This transformation is done by Helioviewer's JP2GEN project\footnote{JP2GEN: \url{https://github.com/Helioviewer-Project/jp2gen}}. The transformation model, as well as their choice of the cut-off points, are primarily based on what the AIA project recommended at the time and how the Helioviewer project team wanted the images to look like. As applying a transform function does not introduce a loss of information in the data, and to ensure that the two sets of distributions are similar in shape, we apply the same data transformation functions that were used in JP2GEN module.\par

\begin{figure}[htp] 
    \subfloat[FITS with max value at 16383\label{fig:sidecut}]{%
        \includegraphics[width=1.0\columnwidth]{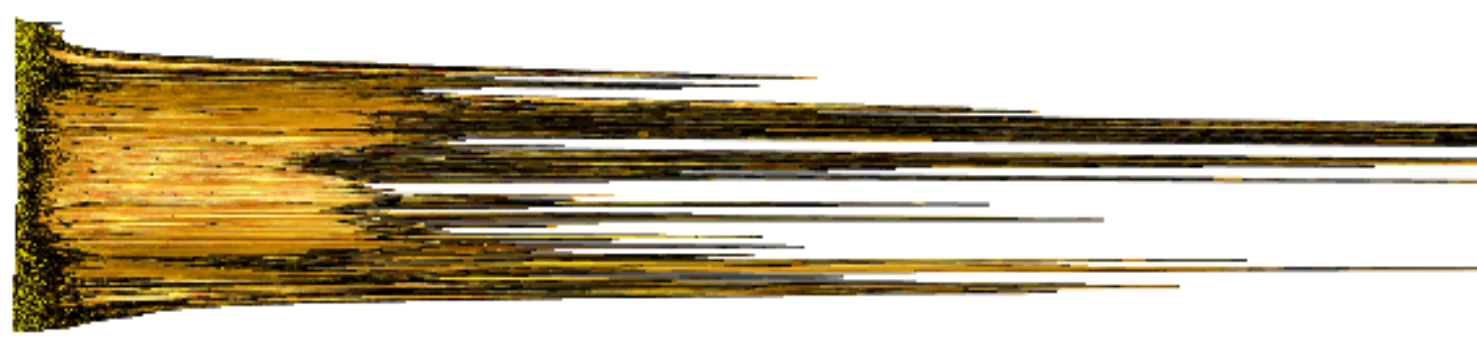}
    }\hfill
    \subfloat[Clipped FITS with max value at 51\label{fig:3dcfits}]{%
        \includegraphics[width=0.48\columnwidth]{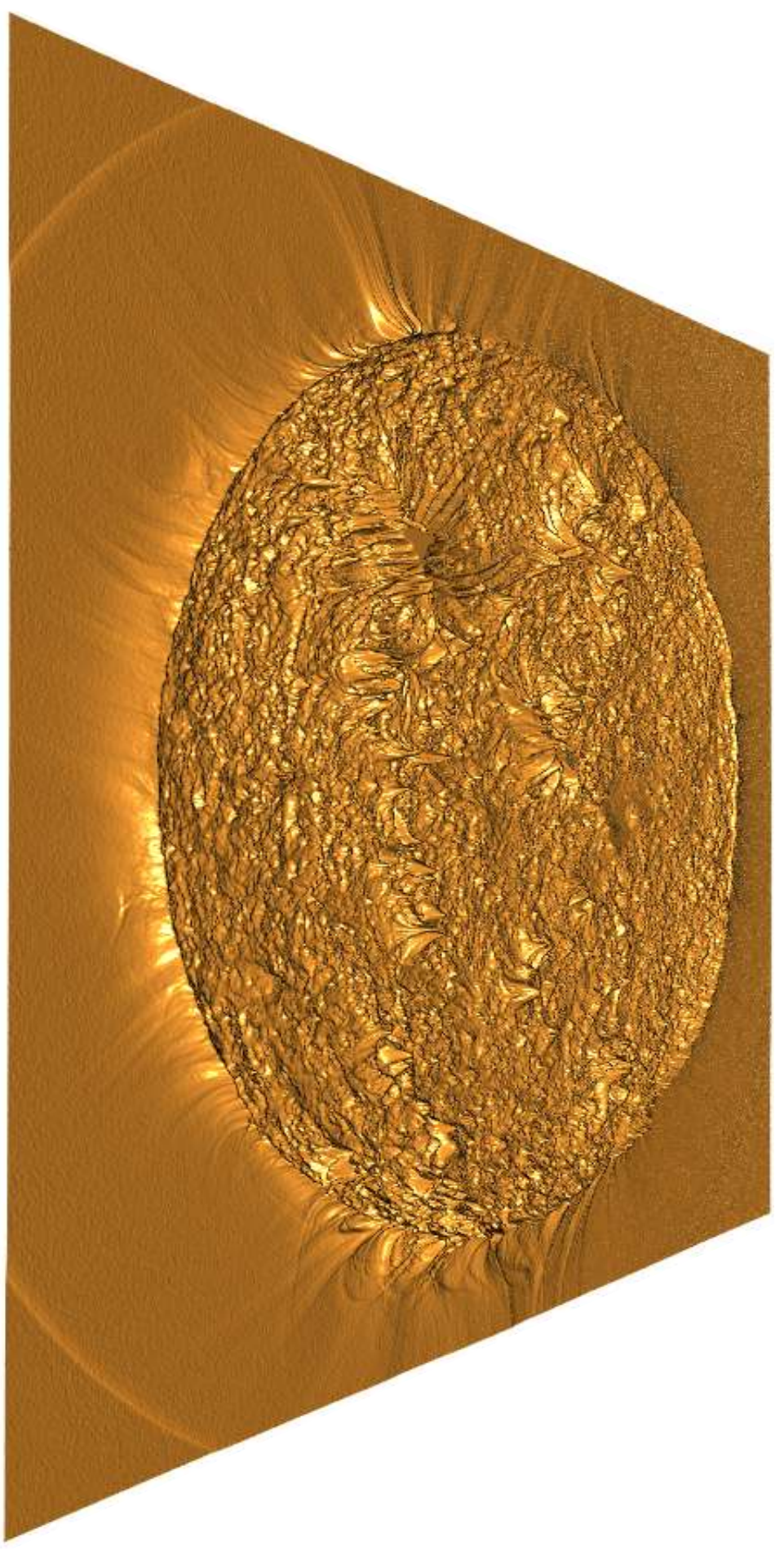}
    }\hfill
    \subfloat[JP2 with max value at 255\label{fig:3djp2}]{%
        \includegraphics[width=0.48\columnwidth]{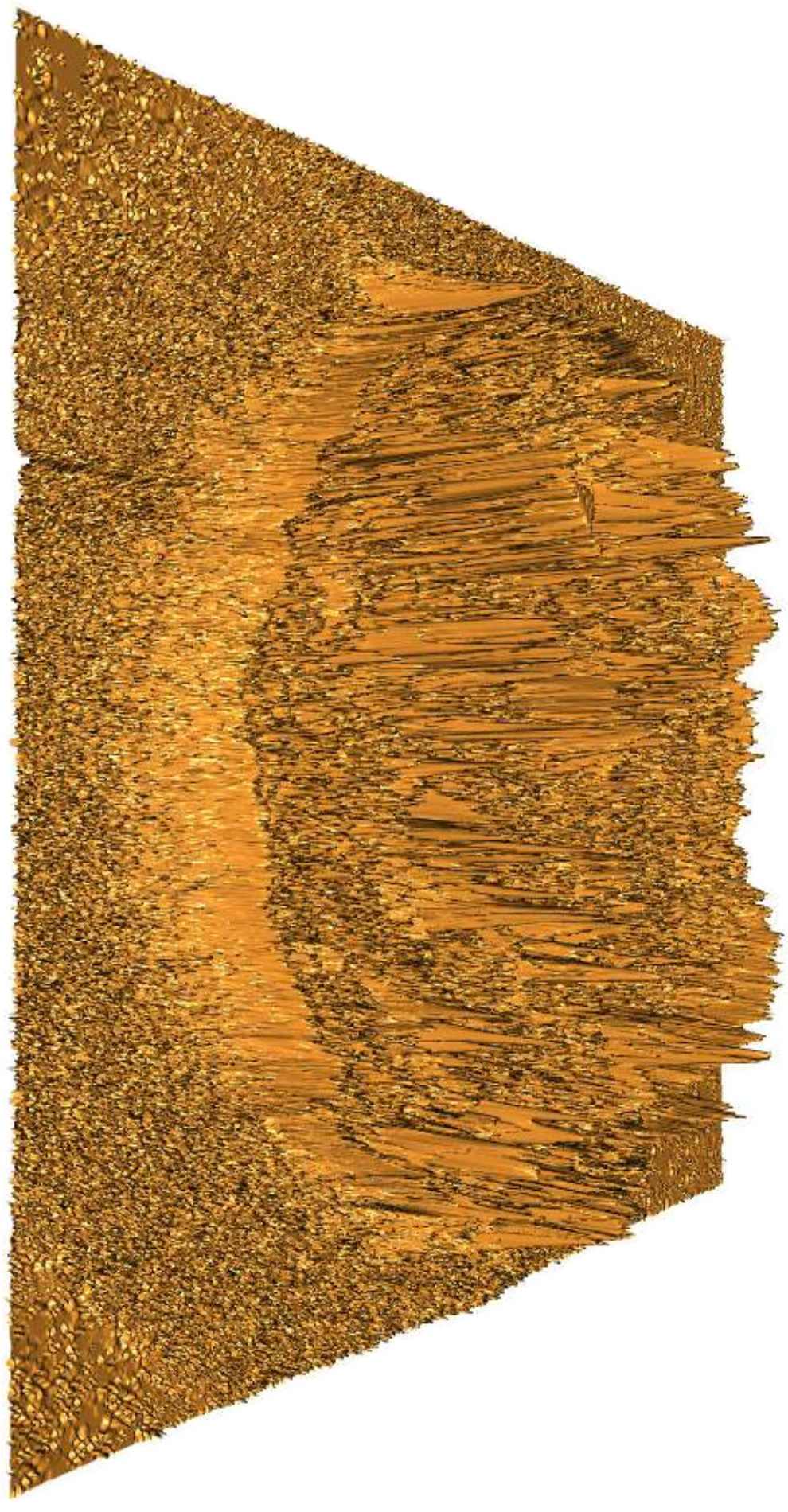}
    }
    \caption{3-D views of an AIA image from the $171$--{\AA} channel, in different formats. The z-axis represents the pixel intensities. Notice that due to the extremely large spikes in the raw FITS image, the full-size of the model for raw FITS image, with the same proportions used in the other two, could not fit here. An un-cut version of this model can be seen in Fig.~\ref{fig:3d}.}
    \label{fig:3dplots}
\end{figure}

The transformation methods differ depending on the wavelength channel of the image. A \textit{linear} transformation is used for $1700$--{\AA} images, a \textit{square root} transformation for images from $171$--{\AA}, and a \textit{logarithm} transformation for the remaining channels. Note that, this is a bijection ($t: \mathbb{N} \longrightarrow \mathbb{R} $) and no data points are removed. The result of such transformation is illustrated in Fig.~\ref{fig:distributions}, on a sample AIA image. It compares the distribution of pixel intensities in a FITS image before clipping (A) and after clipping and transformation (B), and the one derived from the corresponding JP2 image (C). By looking at such comparison, one can see how the hidden bimodal shape of the distribution is perfectly restored after clipping and transformation. This verifies both the correctness and the importance of this step for an unbiased comparison of different image types. In addition to that, a 3D model of the same AIA image in JP2 and in FITS both before and after clipping and transformation is illustrated in Fig.~\ref{fig:3dplots}. In these visualizations, the spikes (representing the magnitude of brightness) reach their highest values at $16383$, $51$ (i.e., $\approx \sqrt{2602}$), and $255$, in FITS, clipped FITS, and JP2, respectively. From this point on we refer to the un-clipped FITS images as \textit{L1.5 FITS}, and to the clipped and transformed FITS as the \textit{clipped FITS}.\par

In this preprocessing step, before clipping of the extreme far-out values, we also take into account the exposure time of the CCD detectors of the AIA instrument for each image. We normalize the pixel intensities based on the specific exposure time with which the image was captured. This is important since it provides us a uniform brightness in our image collection. These values are stored in the header section of each image, under the keyword `EXPTIME', as floating points in double precision (in seconds). \citep{nightingale2011aia}.\par

In summary, we analyze the AIA images in three different formats: L1.5 FITS (as archived in JSOC), clipped FITS, and JP2 (as provided by Helioviewer API) images. The L1.5 FITS and JP2 images are on the two extreme ends of the pre-processing path. L1.5 FITS image are only pre-processed in JSOC for cleaning and calibration in the process of digitizing the images and are relatively large files (varying from $\approx 5$ to $\approx 14$ MB). Whereas, JP2 images are fully pre-processed and compressed (down to $\approx 1$MB) to a typical 8-bit quantized image format. Clipped FITS images lie somewhere in between. They don't have the extreme far-out intensities as the L1.5 FITS images do, but at the same time, they are not limited to $255$ gray levels as JP2 images are. As we mentioned before, we use all these three image types to evaluate our image parameters in Section~\ref{sec:experimental}.\par

\section{Settings of Image Parameters}\label{sec:settings}

Now that we have studied our data types and the image parameters to be tuned, we need to spot the variables in each image parameter that can determine the performance of that parameter. In this section, we provide more information about each of the four image parameters and the implementation details of their computation that allow for the tuning of specific variables and their domains of changes.\par

\subsection{Entropy and Uniformity}

As discussed in Section~\ref{subsubsec:uniformity}, entropy and uniformity parameters solely depend on the normalized histogram of the image color intensities. And it is in the nature of histograms that different choices of the bin size result in different levels of smoothing the histogram. In other words, $p$ in Eq.~\ref{eq:entropy} and \ref{eq:uniformity}, which is the probability density function of the random variable $i$, is defined differently for different bin sizes. Although there are several general rules for determining the bin size, such as Sturges' formula \citep{sturges1926choice} or Scott's rule \citep{scott1979optimal}, often the best choice is the one that is data driven and can be verified by the target classes of the data.\par

So, for these two parameters, the optimal bin size is the variable that will be tuned for utilizing the experiments described in Section~\ref{sec:experimental}. The optimal value of the variable is independently evaluated for each wavelength of image and a set of these values are obtained through the experimental evaluation, one for each wavelength of image we included in the resultant dataset. The domain set for this variable is the set $(0, I) \subset \mathbb{N} \text{ or } \mathbb{R}$, depending on the image type, where $I$ is the max color intensity for the image type under study. For example, the domain set for this variable on the JP2 images from Helieoviewer will be the set of $[0, 255] \in \mathbb{Z}$, whereas the domain set for L1.5 FITS will be the set of $[0, 16383] \in \mathbb{Z}$.\par

\subsection{Fractal Dimension}\label{subsec:regardingFD}

Formerly, in Section~\ref{subsubsec:fractalDimension}, we explained how fractal dimension utilizes the box counting method to measure the dimension of the fractal-like shapes. However, there are a number of different decisions on the implementation of this method that can have an effect on the resultant values that it produces. For instance, the decision on what edge detection algorithm and what values are used for variables of each of the different algorithms will produce differing results. In the following sections, we will explain how this method will be applied to AIA images, and what variables will need tuning in our experimental evaluations of Section~\ref{sec:experimental}.\par

\subsubsection{Box Counting on AIA Images}\label{subsubsec:boxcounting}

To compute fractal dimension image parameter, we first need to know how the box counting method that we discussed before, can be applied on the AIA images. Let us assume that an edge detection algorithm has been chosen and the appropriate settings were found for the algorithm. We can then apply an edge-detection algorithm to an AIA image and then treat the detected edges as the fractals' contour whose dimension is to be measured. Then, for each $\varepsilon$ (box's side length) from a predefined domain, we count the number of grid cells that overlaps with an edge. Considering the resultant pairs, $\langle \varepsilon, N(\varepsilon)\rangle$, as a set of points in the $2$-D feature space of box sizes and the number of boxes, the slope $\beta$ of the fitted regression line can then be measured. $\beta$ is the fractal dimension corresponding to this region. Since the patch size of our image segmentation discussed before is $64\times64$ pixels, the box size in the above procedure will have an upper bound of $64$ pixels. To have a natural sequence of side lengths for these boxes, we use the set of all powers of two within this range, i.e., $\{2, 4, 8, 16, 32, 64\}$, as the domain of the box side length.\par

\begin{figure}[htp]
    \centering
    \includegraphics[width=0.9\columnwidth]{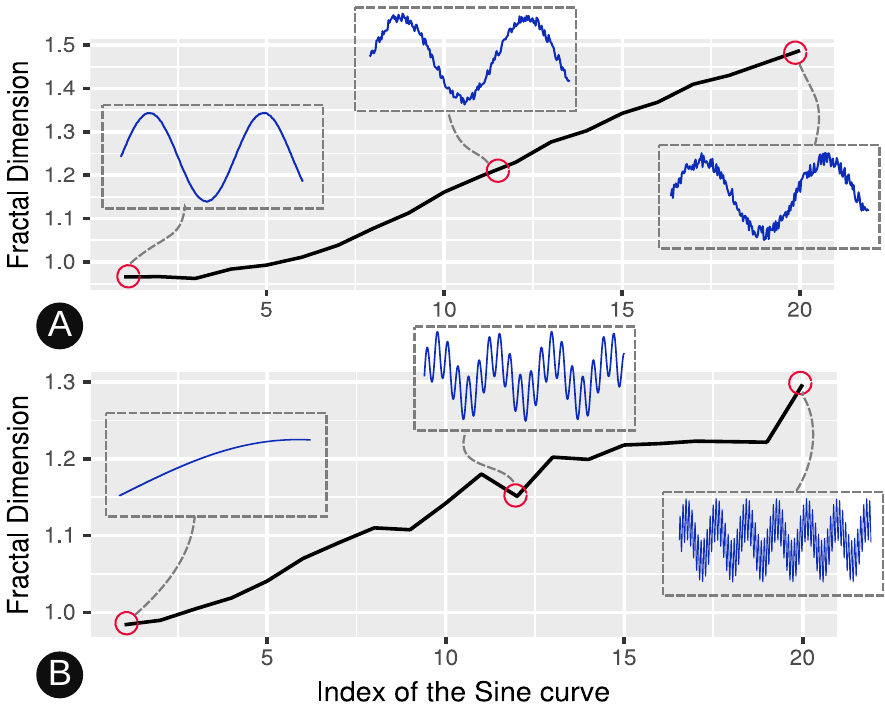}
    \caption{An experiment that shows the growth of fractal dimension on a series of sine waves in two different situations: a) with an iterative increase of random noise to the signal and b) with an iterative increase of frequency of another sine wave to the signal. The results confirms the sensitivity of this parameter to the complexity of the shapes' contour.}
    \label{fig:fdimension}
\end{figure}

Fractal dimension provides a measure to quantify the complexity of the shapes' contour, with larger values indicating higher complexity. In Fig.~\ref{fig:fdimension}, we show how the complexity of a shapes' contour affects the fractal dimension value by using two groups of test signals that are generated to mimic fractal-like shapes. One set is created by adding an incrementally increasing random noise to a sine wave, and the other one, by adding an incrementally increasing frequency of another sine wave to the base sine wave. Measuring the dimension of each signal, a roughly linear growth of fractal dimension is observed that conforms to our expectation.\par

\subsubsection{Edge Detectors}\label{subsubsec:edgedetectors}

The brief explanation of the box counting method tells us that the effectiveness of the fractal dimension parameter in describing the textural feature of an image relies on the quality of the edge detector method that provides the fractal-like shapes. That is, a noisy input, as well as an overly smoothed image, may render this parameter completely ineffective. This fact is the motivation for the following survey of existing edge detection methods and their performance on AIA images. Note that for this application, both the quality of the detected edges that are to be the input to the box counting method, and the execution time of each of the edge detection methods are important, as a longer execution time will require more computational resources for the near real time constraint to be met.\par

Among the existing edge detection methods, we choose Sobel \citep{sobel1990isotropic}, Prewitt \citep{prewitt1970object}, Roberts Cross \citep{roberts1963machine} edge detectors as the classical candidates, Canny's \citep{canny1986computational} edge detector as a popular, modern method, and also SUSAN \citep{smith1997susan} as a less popular but a more recent approach. It has been shown in several different comparative analysis \citep{maini2009study,heath1996comparison,sharifi2002classified} that Canny edge detection algorithm performs better than all of its ancestors in most scenarios, especially on noisy images. Given the special noisy nature of the AIA solar images, with layers of noisy textures instead of solid foreground objects and background landscapes, the classical methods are likely to fail. That being said we do not wish to simply rely on general knowledge about the performance of these methods on textural inputs. Instead, we apply these filters on AIA images and compare the quality of the detected edges that are to be the input to the box counting method.\par

\begin{figure}[htp] 
    \subfloat[A flaring region\label{fig:ceds:original}]{%
        \includegraphics[width=0.48\columnwidth]{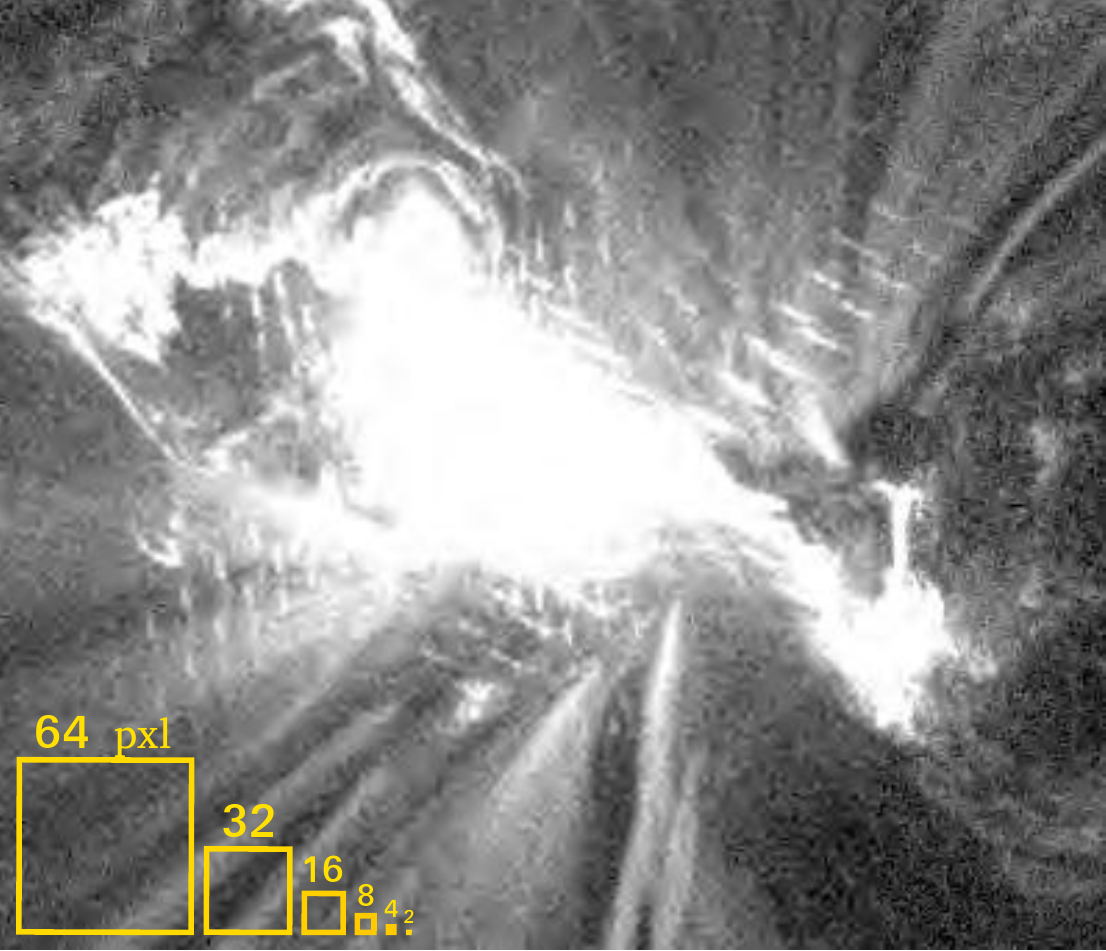}
    }\hfill
    \subfloat[Sobel\label{fig:ceds:sobel}]{%
        \includegraphics[width=0.48\columnwidth]{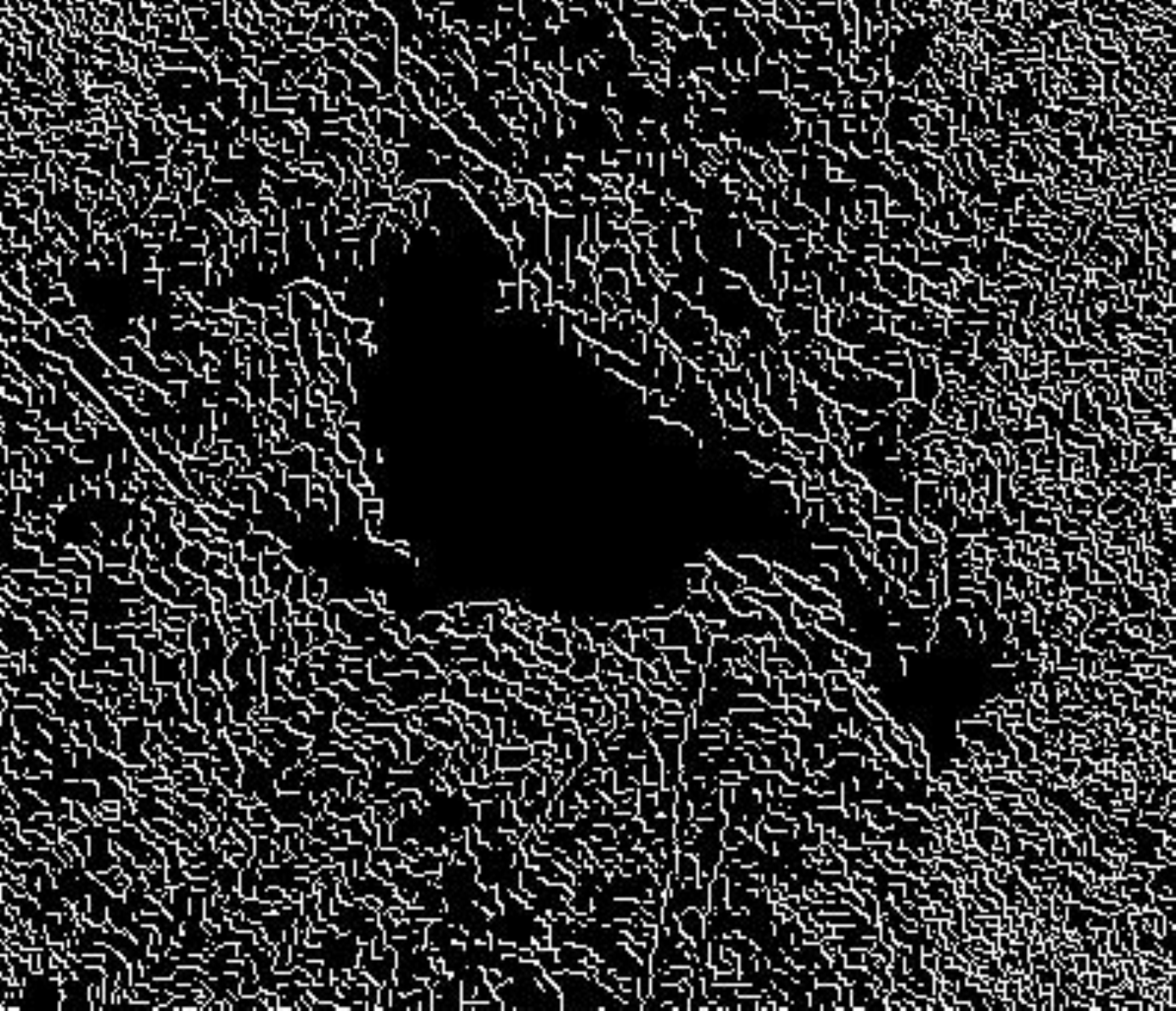}
    }\hfill
    \subfloat[Prewitt\label{fig:ceds:prewitt}]{%
        \includegraphics[width=0.48\columnwidth]{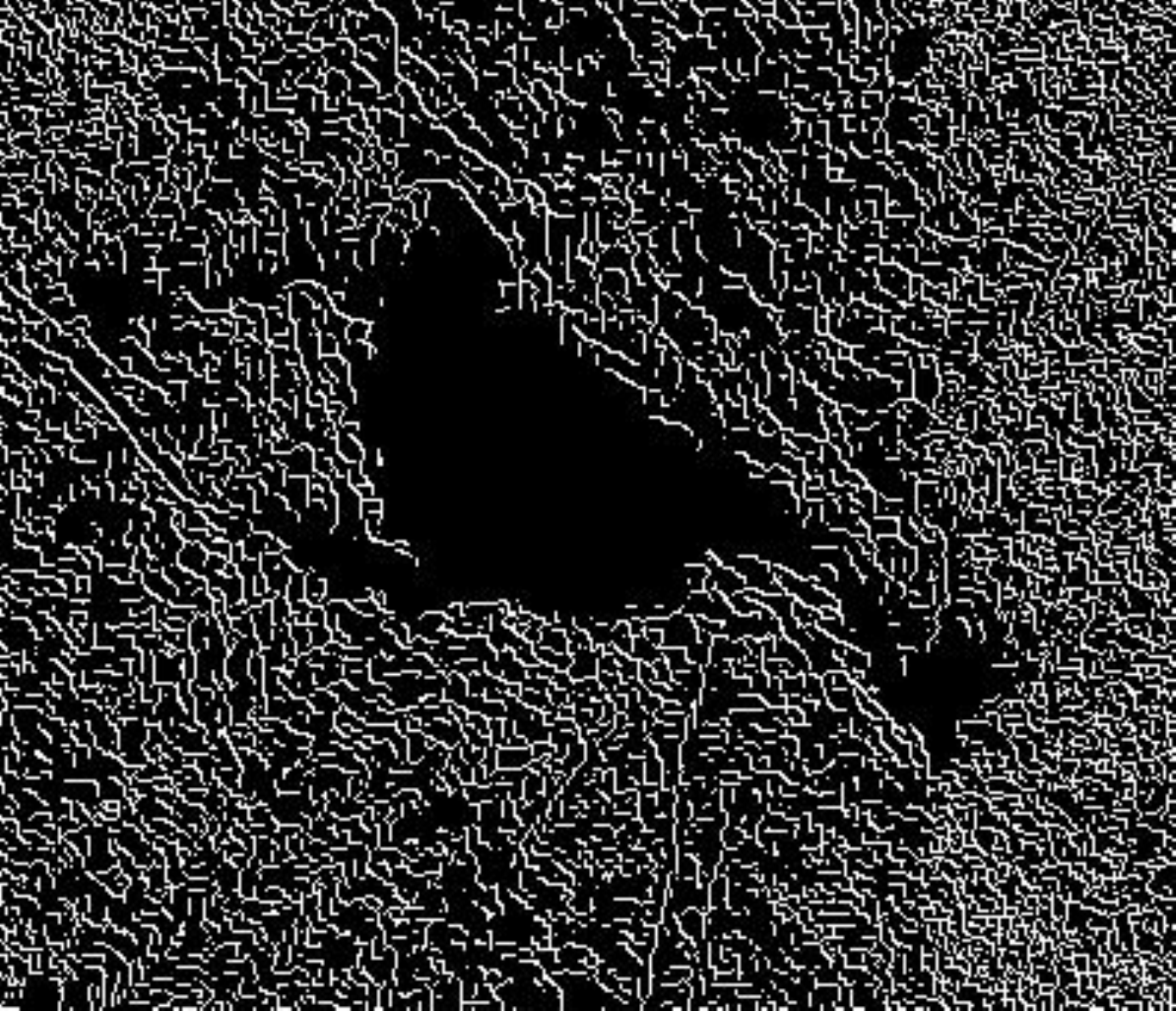}
    }\hfill
    \subfloat[Roberts\label{fig:ceds:roberts}]{%
        \includegraphics[width=0.48\columnwidth]{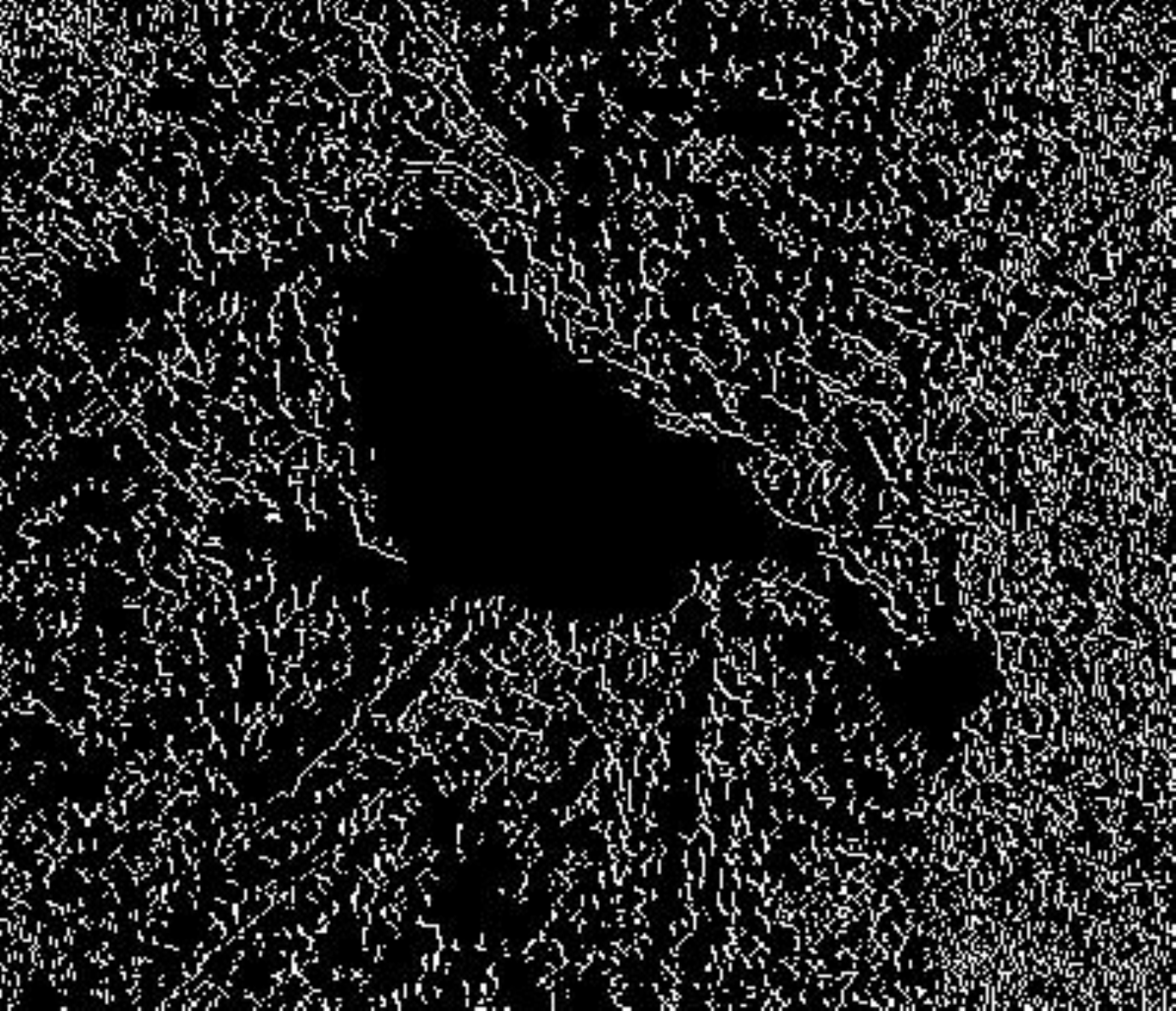}
    }\hfill
    \subfloat[SUSAN\label{fig:ceds:susan}]{%
        \includegraphics[width=0.48\columnwidth]{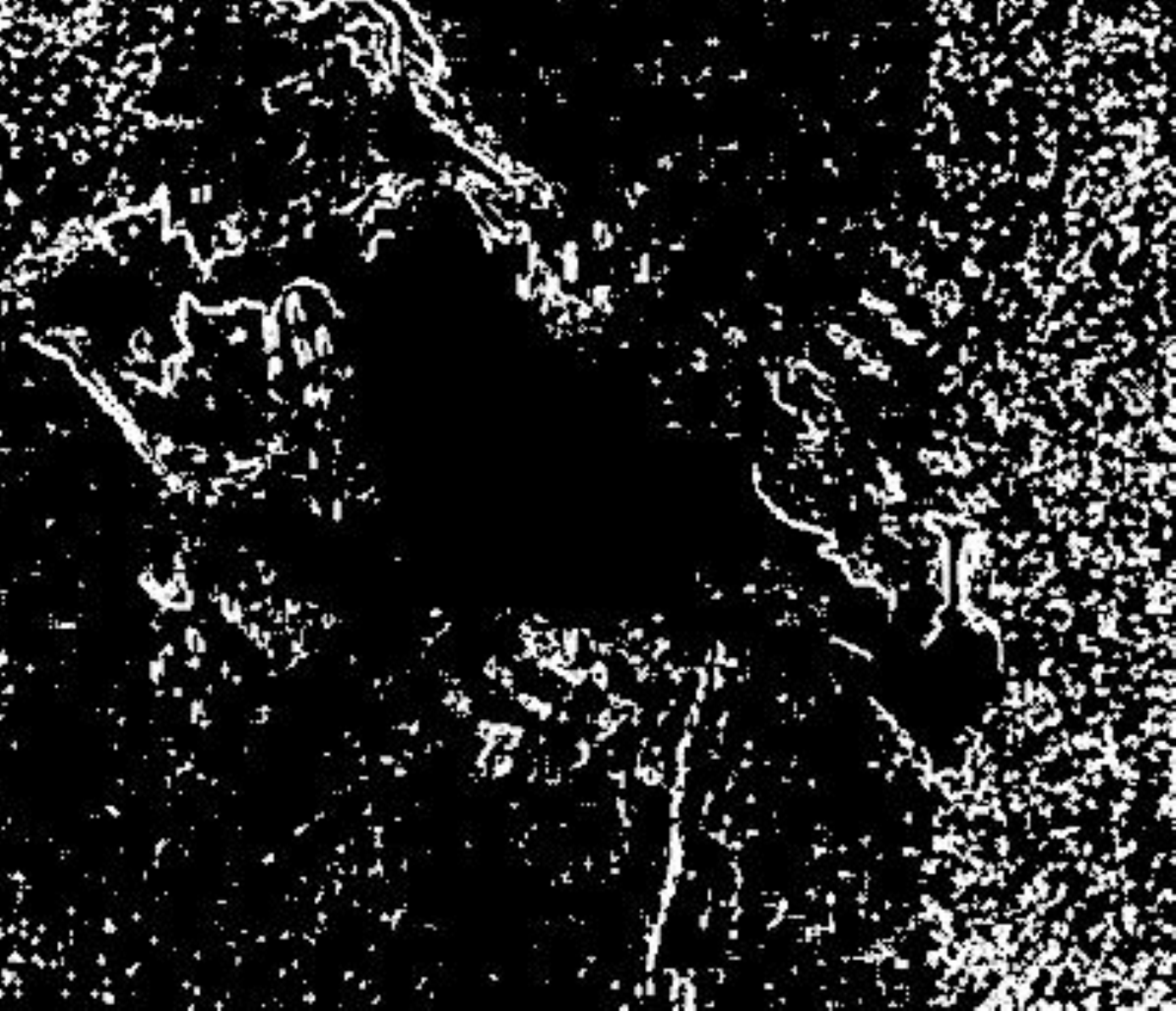}
    }\hfill
    \subfloat[Canny\label{fig:ceds:canny}]{%
        \includegraphics[width=0.48\columnwidth]{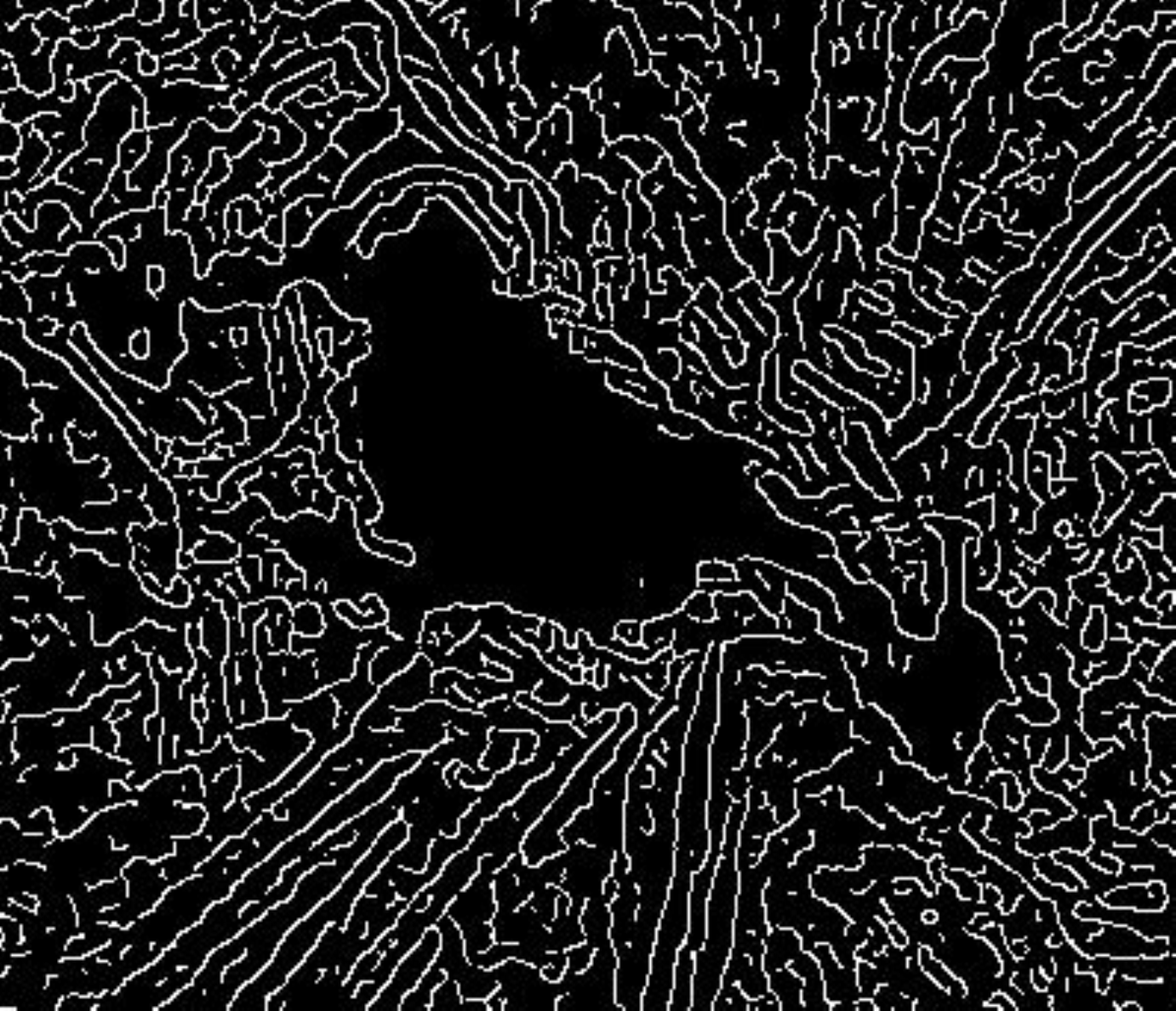}
    }\hfill
    \caption{A cut-out of an active region instance observed on March 7, 2012 at 00:24:14:12 UT from the $171$--{\AA} channel, as well as the outputs of different edge detector methods are shown. In a, the relative size of the boxes (i.e., $64, 32, 16, 8, 4$, and $2$ pixels) used in the box counting method is also illustrated.}
    \label{fig:ceds}
\end{figure}

The first three edge detection methods, Sobel, Prewitt and Roberts Cross, are relatively simple algorithms. They each begin by estimating the first derivative of the image by their corresponding gradient operators (masks). Then, since the magnitude of the gradient vectors do not give thin and clear edges, non-maximum suppression is also applied (as it is done in Canny) to eliminate the multiple representations of each edge. The results of the Sobel, Prewitt, and Roberts Cross methods can be seen in Figures~\ref{fig:ceds:sobel}, \ref{fig:ceds:prewitt}, and \ref{fig:ceds:roberts} respectively.\par

Canny edge detection, however, is more complicated and starts with a prior smoothing step using a $5\times5$ Gaussian kernel. This mitigates the effect of noise on calculation of the gradient. Then, using a $3\times3$ Sobel operator, the gradient of each pixel, $g = (g_x, g_y)$, which is a vector with magnitude $\sqrt{{g_x}^2 + {g_y}^2}$ and orientation $\arctan(g_{y}/g_{x})$, is calculated. Each pixel having nine adjacent neighbors, allows nine different angles for the edge passing through that pixel. Since only the orientation of the edges matter (and not the direction), the choices will be limited to four. Therefore, the continuous range of the calculated angles should be quantized and mapped to one of the following choices: $0^{\circ}$, $45^{\circ}$, $90^{\circ}$, or $135^{\circ}$. This is followed by a thinning process of the edges (i.e., non-maximum suppression) which eliminates the pixels which are labeled as edges but their locations are not in line with the calculated orientation of the edges. At the end, a hysteresis thresholding comes to clean up the disconnectivity of the edges by using two thresholds; a low threshold, $lt$, and a high threshold, $ht$. Any pixel with gradient magnitude greater than $ht$ is labeled as an edge, and a non-edge if its magnitude is less than $lt$. For pixels with magnitudes between $lt$ and $ht$, they are considered part of an edge if and only if they are connected to a pixel which is already labeled as an edge. This last step, next to the initial smoothing step, makes Canny edge detector an expensive filter, but this cost pays off by producing less broken edges and a less noisy output.\par

SUSAN edge detector on the other hand, adopts a very different approach by not using any image derivatives which makes it a good candidate for noisy images like ours. This is the very reason for including it in our list, despite its computation cost. This edge detector has a core concept called Univalue Segment Assimilating Nucleus (in short USAN) which is the central point (nucleus) of the circular masks, and a principle called SUSAN principle which is stated as follows: ``An image processed to give as output inverted USAN area has edges and two dimensional features strongly enhanced, with the two dimensional features more strongly enhanced than edges''. The intensity of the nucleus and the second moment of the area of USAN masks are used to find the edge directions. And eventually, similar to Canny, a non-maximum suppression will be applied to clean up the edges. In this study, we use the implementation of this method from \textit{OpenIMAJ} library \citep{Hare:2011:OIJ:2072298.2072421}.\par

To compare the quality of these edge detectors, we fed each of those methods with a variety of AIA images varying in the queried time of the solar events, wavelength channels, and the appearing event types. Fig.~\ref{fig:ceds} illustrates one of the visual comparisons; a cut-out of an active region instance observed on March 7, 2012 from the $171$--{\AA} channel and the output of each of the above-mentioned edge detectors. As it is visible in this comparison, Canny edge detector provides much cleaner edges and maintains the orientation of the coronal loops (that electrified plasma flows along) of the flaring region, whereas others barely distinguish the texture caused by the powerful magnetic fields from the more quiet (darker) areas. Given that the edges detected are to be passed to the box counting method with the box sizes as large as those shown in Fig.~\ref{fig:ceds:original}, it is visually convincing that for the Sobel-like methods (i.e., Sobel, Prewitt, and Roberts), such a uniform distribution of the extremely short and broken edges does not lead to a reliable measure of the dimension corresponding to different regions. About SUSAN's output (see Fig.~\ref{fig:ceds:susan}), although the results are very different from the others, it does not seem to be a good choice for noisy textures as it does very little in identifying the visible edges.\par

\begin{figure}[htp] 
    \subfloat{%
        \includegraphics[width=0.32\linewidth]{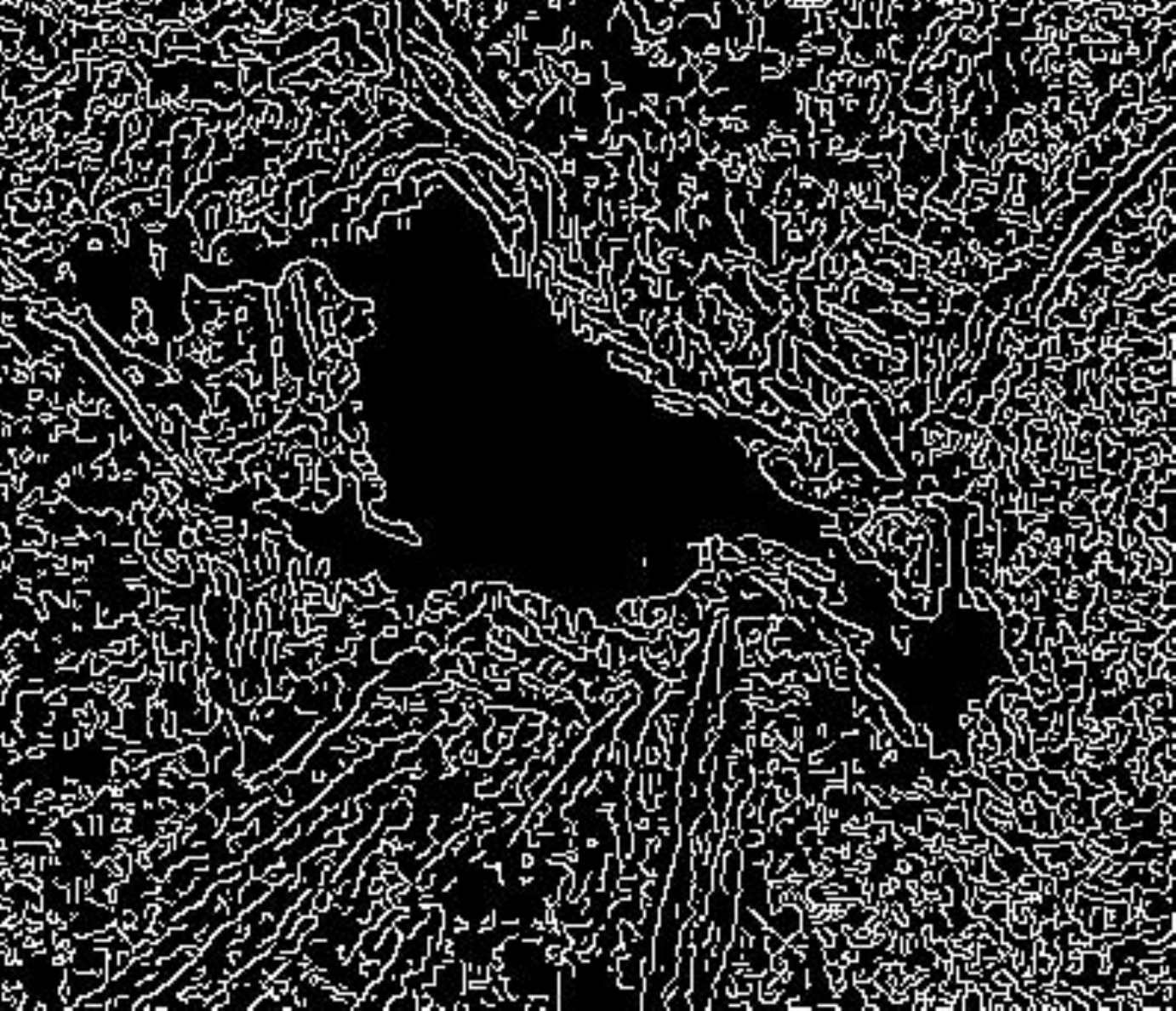}
    }\hfill
    \subfloat{%
        \includegraphics[width=0.32\linewidth]{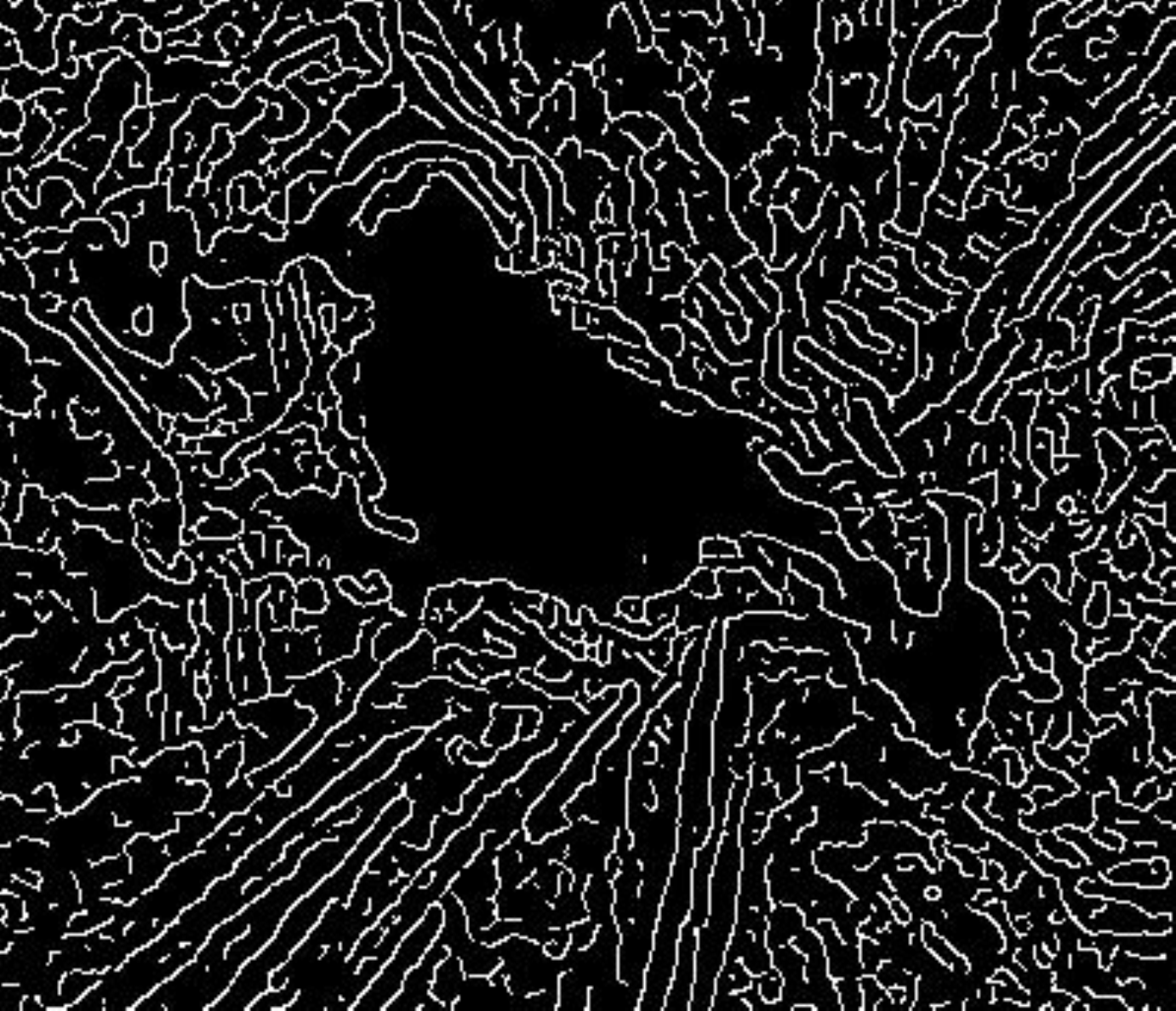}
    }\hfill
    \subfloat{%
        \includegraphics[width=0.32\linewidth]{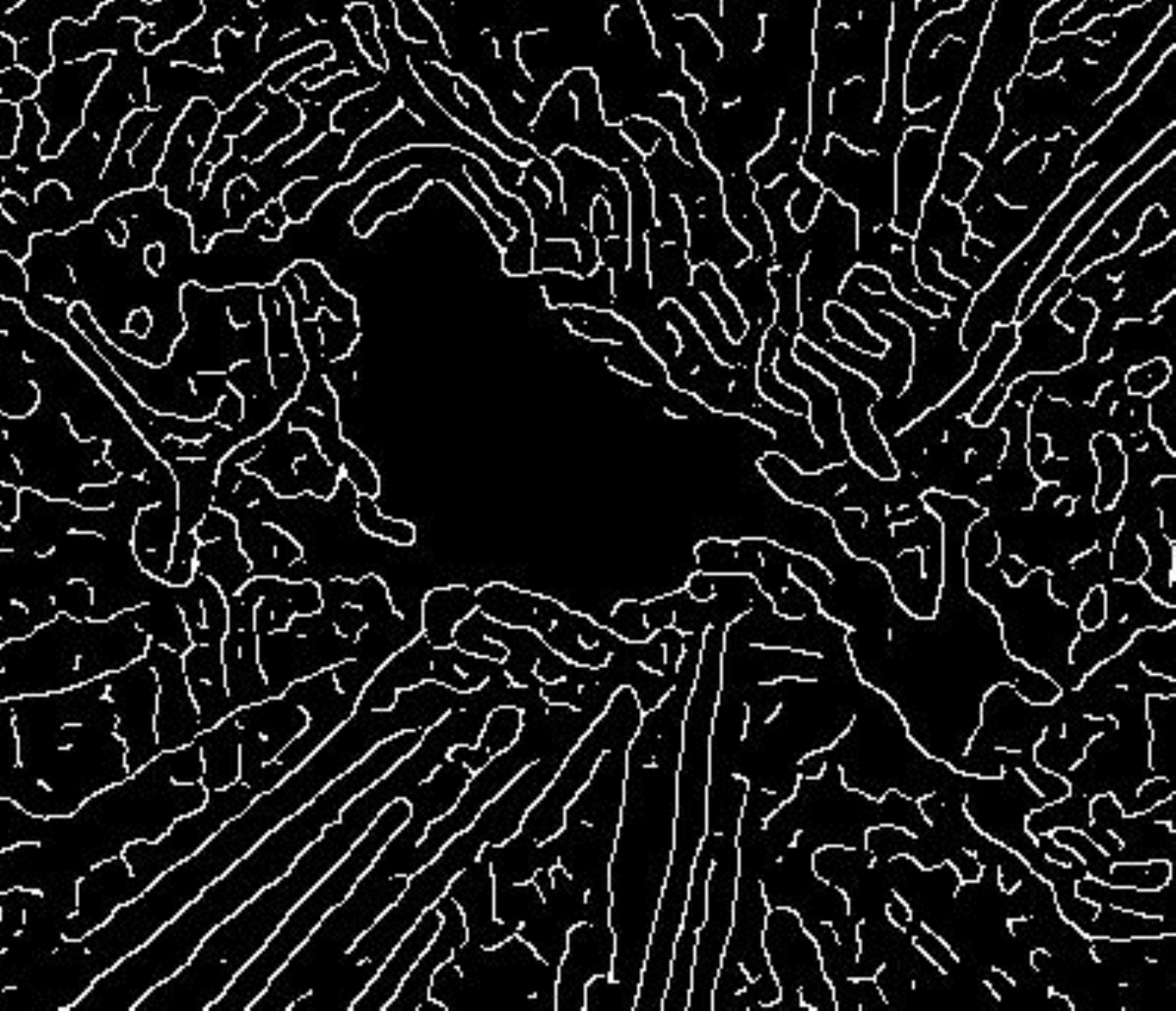}
    }\hfill
    \subfloat{%
        \includegraphics[width=0.32\linewidth]{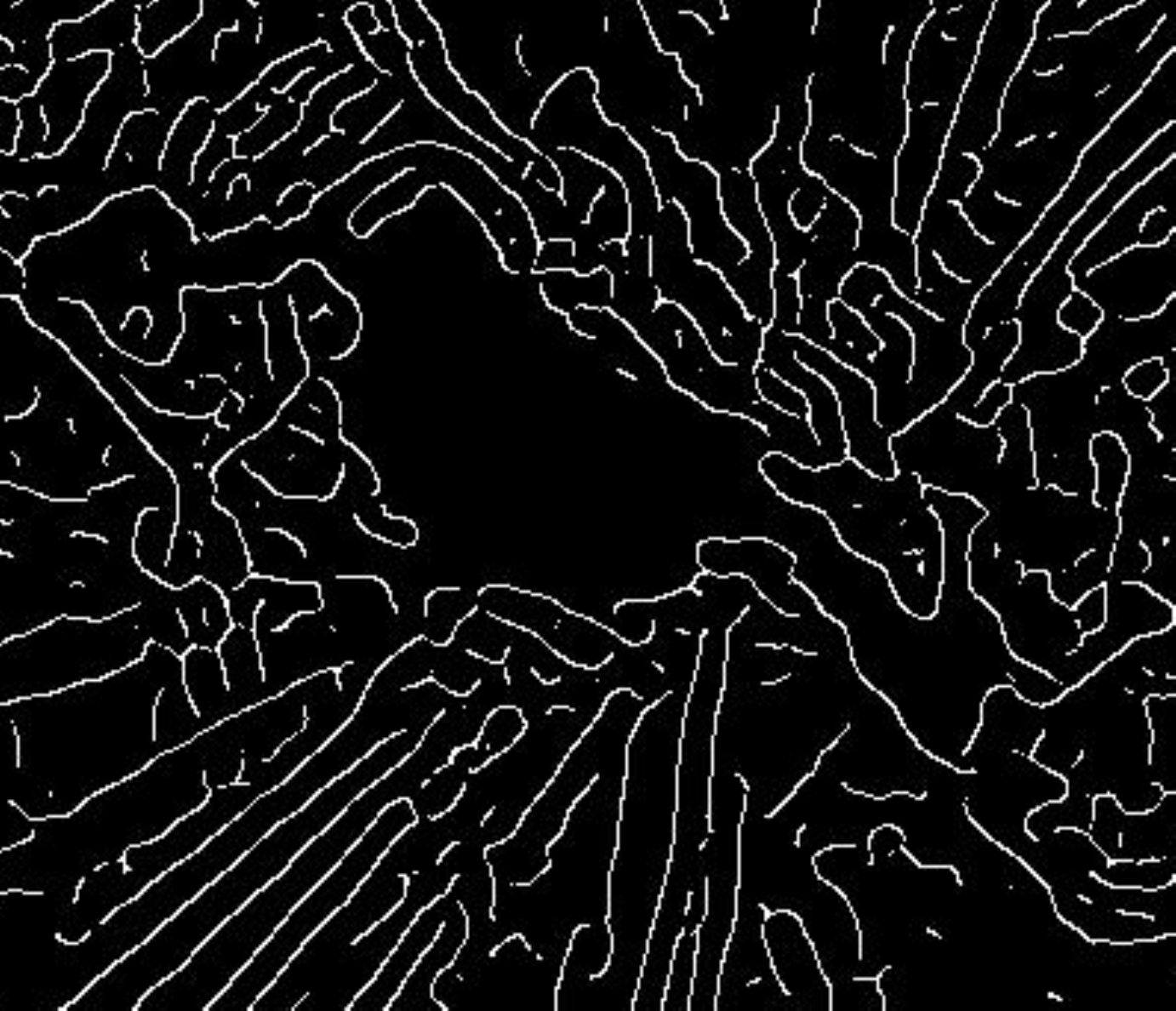}
    }\hfill
    \subfloat{%
        \includegraphics[width=0.32\linewidth]{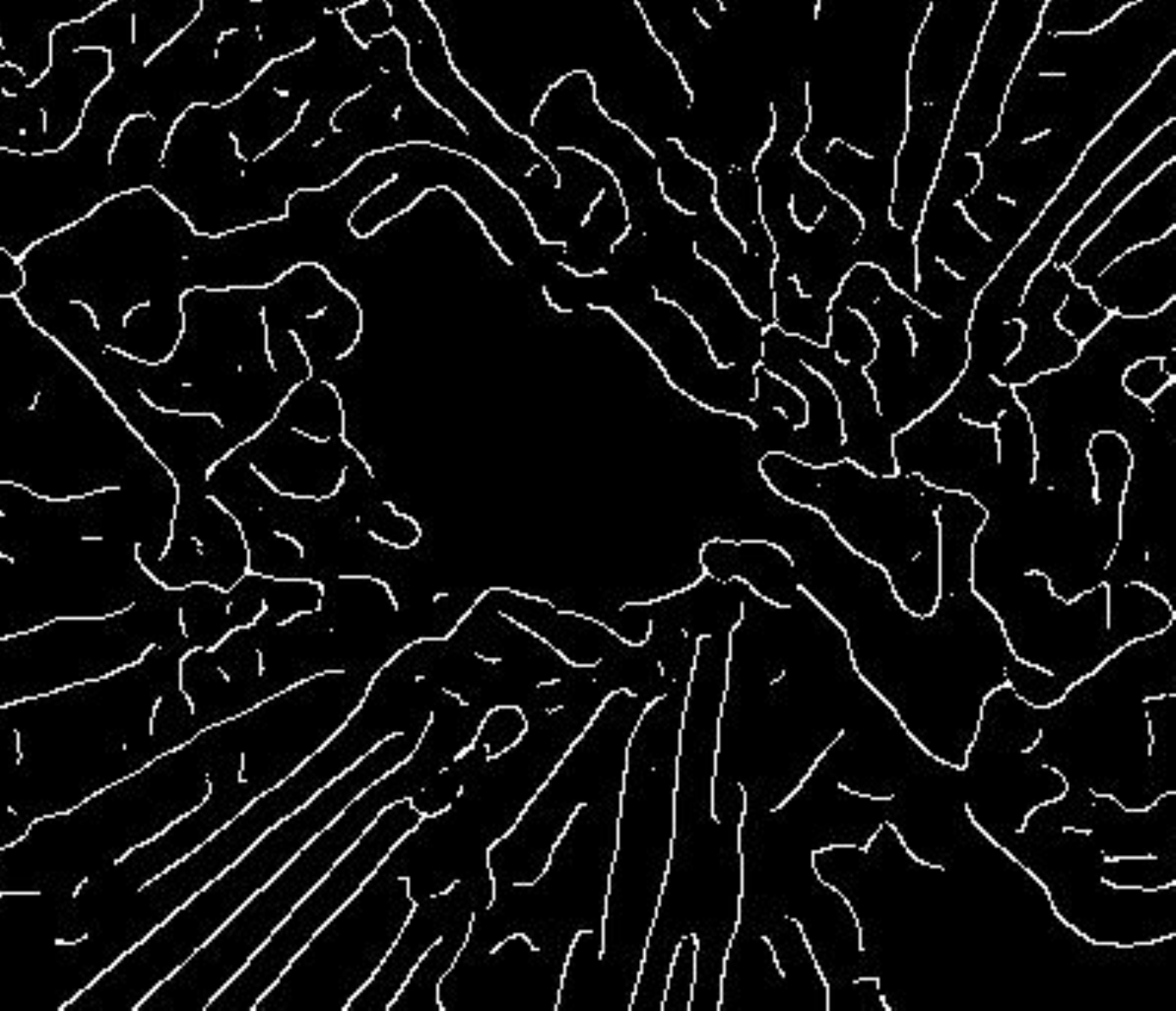}
    }\hfill
    \subfloat{%
        \includegraphics[width=0.32\linewidth]{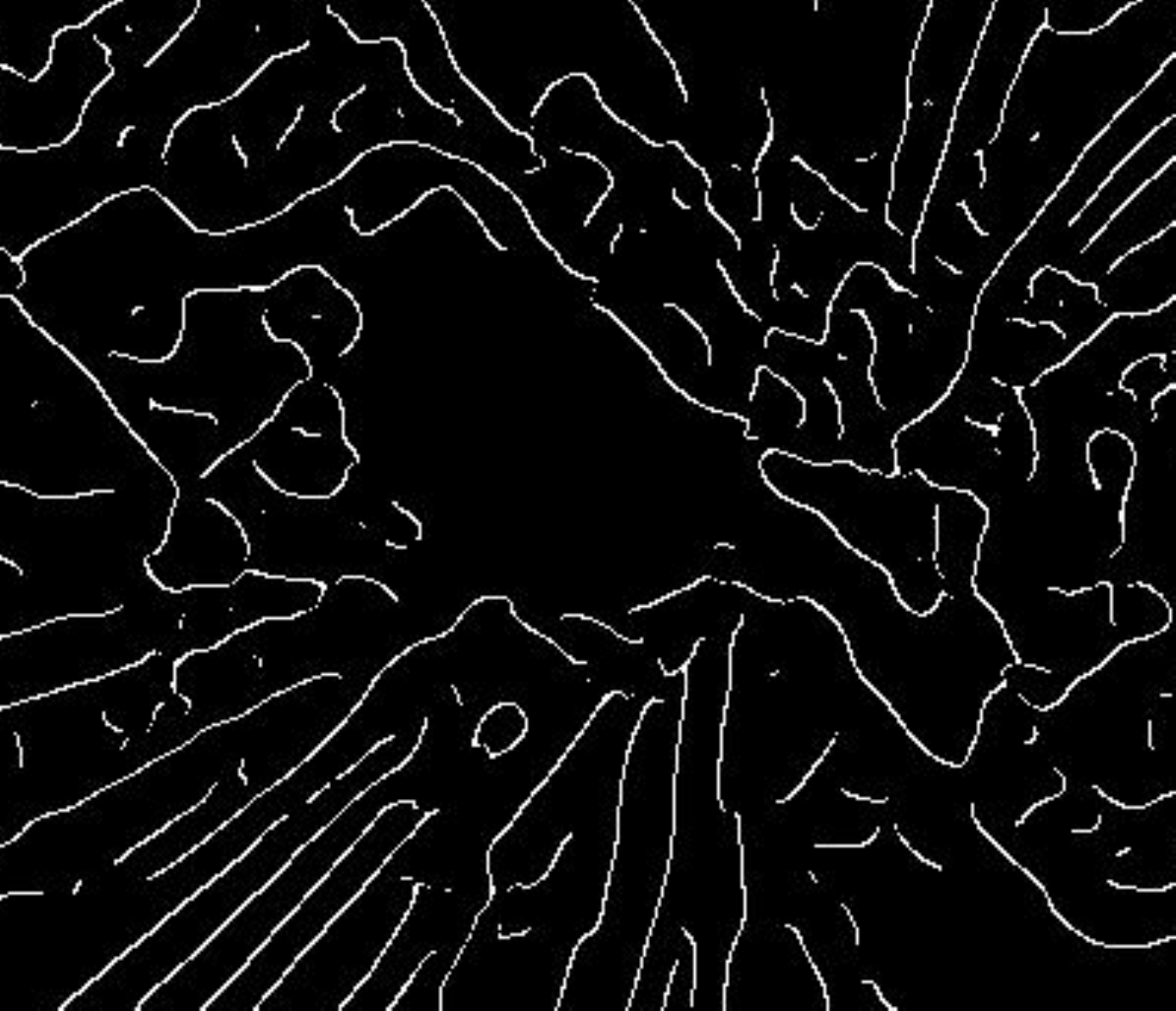}
    }
    \caption{Canny edge detector on an active region instance, with $lt = 0.02$, $ht=0.08$, and $\sigma$ varying from $1$ to $6$, starting at top-left image and ending at the bottom-right.}
    \label{fig:cannys}
\end{figure}

Another argument in favor of Canny edge detector is the tunability of this method that is possible by adjusting its three variables; the standard deviation of the Gaussian smoothing ($\sigma$) and the lower ($lt$) and higher ($ht$) thresholds, as discussed in Section~\ref{subsubsec:edgedetectors}. In Fig.~\ref{fig:cannys}, the effect of such tuning on the same sample active region used before is shown. Note the smooth decrease in the noise level as $\sigma$ increases while the general patterns and directions are maintained.\par

Regarding the running time of these methods, Table~\ref{table:times} summarizes our comparisons. Although, the execution time of the utilized methods is an important factor in general, in this case, it does not seem that there are many choices left for us, except the relatively most expensive one, i.e., Canny edge detector. This is because only this method is producing the relevant input for the box counting method of the fractal dimension parameter. The decision is between a faster method which mostly produces uniform noise, and a relatively more expensive one that provides the right input (where the physical characteristics such as the coronal loops as the curving lines of powerful magnetic fields are enhanced) for fractal dimension.\par

The results listed in Table~\ref{table:times} are the average execution time measured by running each of the algorithms on a group of $100$ full-disk AIA images of size $4096\times4096$ pixels in $10$ different wavelength channels, having different event types. To put the numbers in context, it is worth noting that these experiments are conducted on a Linux machine with a core $i5-6200U$ CPU, $2.30$GHz$\times4$, and an $8$GB of memory, while for any operational task, a much more powerful machine would be used to process the images. Therefore, the running time of Canny edge detector is expected to be less than $3.619$ seconds for a single image.\par

\begin{table}
    \centering
    \caption{The average execution time for different edge detection methods on $4096\times4096$--pixel AIA images.}
    \begin{tabular}{c l r}
          & Method      & Execution Time (Sec.)\\ \toprule
        1 & Sobel       & $2.267$ \\ \midrule
        2 & Prewitt     & $2.208$ \\ \midrule
        3 & Roberts     & $1.809$ \\ \midrule
        4 & SUSAN       & $0.674$ \\ \midrule
        5 & Canny       & $3.619$ \\ \midrule
        \bottomrule
    \end{tabular}
    \label{table:times}
\end{table}

Having Canny edge detector chosen as the method to filter the input AIA images and pass them to the box counting method, tuning of fractal dimension parameter would then depend on the choices of $lt$, $ht$, and $\sigma$ of the edge detector. Our experiments show that by changing $\sigma$ while having $lt$ and $ht$ fixed at a narrow interval close to zero (e.g., $lt = 0.02$ and $ht = 0.08$), we could cover almost the entire spectrum of the possible outputs. This observation leaves only one variable, $\sigma$, for the tuning of this image parameter.\par

\subsection{Tamura Directionality}\label{subsec:regardingDirectionality}

The general formula to compute the directionality parameter was explained in Section~\ref{subsubsec:tamuraDirectionality}. As it calculates the weighted variance of the gradient angles, it requires the gradient of the image to be calculated beforehand. For an image $I$, the gradient vector is:
\begin{equation}
    \nabla I = [g_x, g_y] = \Bigg[ \frac{\partial I}{\partial x}, \frac{\partial I}{\partial y}\Bigg]
\end{equation}
from which the direction and magnitude of the vectors can be calculated as follows:
\begin{equation}
    [\phi, r] = \big[ \arctan \big(\frac{g_y}{g_x}\big), \sqrt{ {g_x}^2 + {g_y}^2}\big]
\end{equation}

There are different kernel convolution matrices used to approximate the gradient vector of an image. Since no preprocessing such as smoothing is required for this task, their computation time depends only on the kernel size. Therefore, we limit our choices to the simple but well-known gradients, such as Sobel--Feldman \citep{sobel1990isotropic}, Prewitt \citep{prewitt1970object}, and Roberts Cross \citep{roberts1963machine}. The last one has a $2\times2$ kernel matrix that makes it slightly faster but more sensitive to noise, due to its smaller kernel matrix comparing to the $3\times3$ matrices of the other two. After we visually studied the remaining two kernels, we observed that both their gradient outputs and the histograms of angles are fairly similar. Therefore, we decided to utilize Sobel--Feldman as our gradient mask, which seems to be more popular and widely used in different libraries and applications.\par

From the derived gradient matrix, the histogram of angles can be computed and passed to Eq.~\ref{eq:tdir}. Now, the tuning of $T_{dir}$ has come down to a peak detection method that identifies the ``dominant'' peaks. Therefore, to achieve any improvement on this parameter, a peak detection algorithm must be utilized. There has been a great deal of effort in identification of peaks and valleys, specially in the domain of time series analysis and signal processing \citep{palshikar2009simple}. But it is important to note that peak identification is a subjective task that is often determined by the general behaviour of the data under study. Since peak detection tends to be a domain specific task, where each domain has different criteria for the definition of peaks, it is logical to design a peak detection method which is more compatible with the type of the data we have, i.e, the distribution of the gradient angles of the AIA images. The method that we have chosen to utilize is explained in greater detail in \citep{ahmadzadeh2017improving}. In the next section, we briefly review this approach.\par

\subsubsection{Peak Detection}

In general, the peak identification task is to determine the domains, $d_i$, within which the local maxima of the data sequence $C = \{c_1, c_2, \cdots, c_n\}$ are located. In other words, the goal is to identify $d_i$'s such that $\exists c_i\in d_i, \forall c \in d_i, c_i \geq c$. We build our algorithm on the basis of a na{\"i}ve assumption that it is enough for each data point to be compared only with its adjacent points in the sequence, meaning that for a local maximum $c_i$, the domain would be $d_i = \{c_{i-1}, c_i, c_{i+1}\}$. If $c_i$ satisfies the condition, we consider it a \textit{candidate peak}. Then we pass the candidate peaks to a three--fold filtering process to pick only the most significant ones. At each step, we check one of the user--defined criteria, namely the threshold, $t$, the minimum distance, $d$, and the maximum number of peaks, $n$. First, we remove all candidate peaks which lie below the threshold $t$. The peaks which are too close to a dominant one will be removed in the next step. Starting from the identified peaks with greater values we simply remove their neighbors within the radius of $d$. And finally, just to provide a control tool for the cases where a certain count of the peaks is of interest, we keep the top $n$ peaks and drop the rest.\par

The proposed algorithm, in spite of its simplicity, provides a flexible tool to determine the significance of the dominant peaks in a data--driven fashion. Using this algorithm, tuning of this parameter is bound to the three above-mentioned variables of the peak detection method.\par

\subsection{Summary of Settings}\label{subsec:settingsSummary}

In Summary, for each image parameter we managed to identify the variables and their domains, that play a role in tuning of that parameter. We use these variables to find the best settings for the image parameters to obtain the highest accuracy in prediction of the solar events. The variables of interest for each of the four image parameters are summarized below:

\begin{enumerate}
    \item Uniformity: the number of bins, $n$,
    \item Entropy: the number of bins, $n$,
    \item Fractal dimension: the Gaussian smoothing parameter used in Canny edge detector, $\sigma$,
    \item Tamura directionality: the threshold, $t$, the minimum distance, $d$, and the maximum number of peaks, $n$, used in our peak detection method.
\end{enumerate}

\section{Experimental Analysis}\label{sec:experimental}

In this section we discuss the tuning process of the image parameters listed in Table~\ref{table:tenParams}. We start with explaining our methodology as our general approach towards tuning the parameters, and then we elaborate on the details of the task for each of the four image parameters separately. Finally, we report the performance of each of the parameters in classification of active region, coronal hole and quiet sun event instances.\par

\subsection{Methodology}\label{subsec:methodology}

Among the ten image parameters, the descriptive statistics (i.e., $\mu$, $\sigma$, $\mu_3$, $\mu_4$) depend only on the intensity value of the pixels. On the basis of these statistics, relative smoothness and Tamura contrast can be then calculated. None of these six parameters have any constraints, thus not tunable. For the remaining four parameters, we run a univariate parameter tuning process on their constraints which we identified in Section~\ref{sec:settings}.\par

For each parameter, first, we find the set of $n$ key constraints (or variables), and identify appropriate numeric domains, $d_i$, for each constraint $i \in \{1,2, \dots, n\}$. As a result, we will have a feature space of dimension $|d_1| \times |d_2| \times \dots \times |d_n|$, for that particular image parameter, where $|d_i|$ is the cardinality of the domain set $d_i$. In addition, to describe a particular event, a region of interest must be processed that spans over a variable number of grid cells. This presents the problem of comparing variable sized regions of interest in order to find the optimal setting for the various parameter variables. For instance, if the region spans over $k$ grid cells, it will then be represented by a vector of length $k$, for each image parameter.\par

So, in order to compare the variable sized regions of interest that produce different-length vectors, we use a seven-number statistical summary on the resultant vectors. This process will map each variable sized parameter vector that is computed on a region to a consistent length vector of seven different values. These vectors are computed independently for each of the $9$ ultraviolet (UV) and extreme ultraviolet (EUV) wavelength channels from the AIA that we include in our investigations. Since these channels produce significantly different images of the Sun, we expect that each channel will require individual tuning of the parameter calculation variables in order to take such differences into consideration and produce the best results for each wavelength.\par

Clearly, even for a very small domain for the constraints of any one parameter, a high-dimensional space will be generated by this statistical summary method and therefore, dimensionality reduction is necessary to minimize the effect of the well-known curse of dimensionality. To this end, we use the F-test statistic to rank each of the settings and then select the best ones per wavelength. We use only the best settings to produce our final feature space, which is then utilized to provide a comparison of the three different input image types through a supervised classification of solar events. The ranking process in F-test relies on grouping of the data and measuring the ratio of between-group variability and within-group variability.\par

Our methodology can be summarized in the following five steps:
\begin{enumerate}
    \item Determining the dimension of the feature space (i.e., identifying the constraints and their domains),
    \item Building the feature space for the period of one month (i.e., January 2012),
    \item Reducing the dimensionality of the feature space using F-test (i.e., finding the best settings per wavelength),
    \item Building the (reduced) feature space for the period of one year (i.e., 2012),
    \item Measuring the quality of the parameter using supervised learning.
\end{enumerate}
In the following sections, after we talk about the dataset we used for our experiments, we explain the specific details of our methodology for each parameter.\par

\subsection{Dataset for Supervised Learning}

For the learning and classification phase, we employed the same methodology in collection of data that was used by \citet{schuh2017region} to collect one year worth of AIA images over the entire $2012$ calendar year and the spatiotemporal data related to the solar events reported in this period. Here, we only briefly explain the data acquisition process and refer the interested reader to the article where the entire process is explained in great detail.\par

We target two solar event types, namely active region (AR) and coronal hole (CH), which are in particular of interest for heliophysicists and also because of their similar reporting characteristics that make region identification easier.
As our ground truth, we rely on the AR and CH catalogs of the HEK (Heliophysics Events Knowledgebase) which are detected by SPoCA (Spatial Possibilistic Clustering  Algorithm) \citep{hurlburt2010heliophysics}. In year $2012$, HEK reported $13,518$ AR and $10,780$ CH event instances, at approximately a four hour cadence. Since there are more AR instances, we first collect all of those instances and then we look for CH instances within a time window of $\pm 60$ minutes from each report of an AR instance. Those AR instances that could not be paired with a temporally close CH instance are dropped. The report of each event contains both temporal and spatial information. We use the time stamps of the reports to retrieve the corresponding AIA images (in JP2 and FITS format). The spatial data of each instance consists of a center point for the reported event, its bounding box, and polygonal outline. We use the bounding boxes to extract the image parameters on the region corresponding to each event instance in our training and test phase. With such constraints, we managed to retrieve $2,116$ unique pairs of AR and CH instances. As our supervised learning model requires a control class, an event type that points to a region of solar disk with no report of any other solar events, an artificial event called quiet sun (QS) is introduced. To collect a set of such instances temporally linked to our AR-CH collection, for each report of an AR event, the bounding box of that event is used to randomly search for regions that have no intersection with any reports of AR or CH events.\par

\subsection{Determining the Feature Space}\label{subsec:determining}

Generally, in the machine learning discipline, a feature is a measurable property of a data point being observed. For instance, for AIA images as the data points in our study, entropy of the pixel intensities of an image is a feature derived from that image. Given $d$ different features, a feature space, is a $d$-dimensional space where each of its dimensions corresponds to one of the features. Here, we are trying to tune our image parameters one by one, and we may have one or more variables for each image parameter. So, instead of having multiple features, we are dealing with multiple variations of a single feature. In other words, we derive multiple features from one single parameter and consider them as different features. Therefore, the feature space defined by an image parameter with one variable that takes $|d|$ different values, is a $d$-dimensional space. Similarly, for an image parameter with two variables, a ($|d_1| \times |d_2|$)-dimensional space will be generated, where $|d_i|$ is the cardinality of the domain set for the $i$-th variable.\par

\subsubsection{Feature Space for Entropy and Uniformity}

The admissible feature space suggested by entropy or uniformity parameter is a $d$-dimensional space, where $d$ is the cardinality of the candidate set for the number of bins. The evaluation of both entropy and uniformity is therefore defined as a search over a uniformly distributed number of bins to find the best performing set for our classification task. For the original images in both JP2 and FITS format, the pixel intensities vary within a fixed range, and therefore, the general form of the candidate set can be formulated by the following formula:
\[\Bigl\{k \cdot {\Bigl\lfloor \frac{max - min}{l} \Bigr\rfloor}; \quad l \in  \mathbb{N}, k \in \{1,2,3, \cdots, l\} \Bigr\}  \]
where $l$ is the bin size, and $k$ is a scalar.\par

For JP2 images ($min = 0$, $max = 255$), our visual experiments show that $l = 20$, letting the number of bins be chosen from the set $\mathcal{N}_{JP2} = \{12, 24, 36, \cdots, 255\}$, gives us a comprehensive enough candidate set for creating the feature space. Using such a set, $21$ different entropy (similarly uniformity) parameters will be generated, with bin widths ranging from $1$ to $21$ units. Similarly, for L1.5 FITS images, ($min = 0$, $max = 16383$), the number of bins will be chosen from the candidate set $\mathcal{N}_{FITS} = \{780, 1560, 2340,\cdots, 16383\}$.\par

For the clipped FITS images, however, since the $max$ values differ from one wavelength to another, the candidate set should also adapt to the corresponding range. As the new maxima are much smaller than the global maximum, due to the transformation of the pixel values (discussed in Section~\ref{subsubsec:transformation}), the above model results in bagging of most of the pixel intensities in one single bin and leaving the other bins empty. To avoid such an overly smoothed histogram, in addition to substituting the after-clipping maxima instead of the global maximum, we downsize the bins by a factor of $10$. This is of course meaningful since for the clipped images, the pixel intensities are real numbers, as opposed to the integer intensities in the L1.5 FITS images. For example, for AIA images from $94$--{\AA} channel, since the after-clipping range of the pixel intensities is $[0, 44]$, the candidate set for the number of bins would be $\{20, 41, 62, \cdots, 440\}$, where in the most extreme case, the bin size will be as small as one tenth of a unit (i.e., $440$ bins for the interval $0$ to $44$). In general, regardless of the wavelength, $|\mathcal{N}_{JP2}| = |\mathcal{N}_{FITS}| = |\mathcal{N}_{cFITS}| = 21$.\par

\subsubsection{Feature Space for Fractal Dimension}
Our experiments in Section~\ref{subsec:regardingFD} concludes that the feature space formed by this image parameter will be determined only by the domain of the variable $\sigma$ in Canny edge detection method. They also show that for $\sigma$ greater than $5$ (when $lt = 0.02$ and $ht = 0.08$) the results are very similar to one another and they all maintain only the very strong edges. Observing the amount of changes in the output as $\sigma$ increases, suggests that the candidate set $\mathcal{S} = \{0.0, 0.5, 1.0, \cdots, 5.0\}$ generates an admissible space.\par

\subsubsection{Feature Space for Tamura Directionality}

As our analysis in Section~\ref{subsec:regardingDirectionality} shows, the variables in our peak detection method, i.e., $t$ and $d$, determine the feature space for Tamura directionality. As for the threshold on the frequency domain of the peak detection method, we consider the first, second, and third quartiles of the frequency, below which the peaks would be ignored, as our candidates. We also add the $90$-th percentile to allow observing the results for the cases that only the significantly dominant peaks are to be taken into account. The domain for this variable is therefore the set $\mathcal{T} = \{0.25, 0.50, 0.75, 0.90\}$.\par

\begin{figure}[t] 
    \subfloat[JP2]{%
        \includegraphics[width=0.23\linewidth]{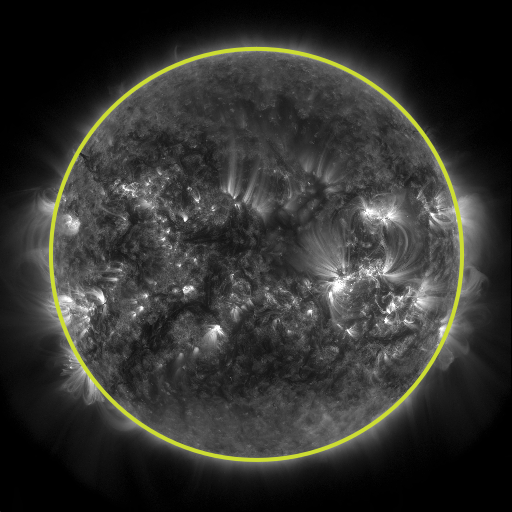}
    }\hfill
    \subfloat[$d = 1$]{%
        \includegraphics[width=0.23\linewidth]{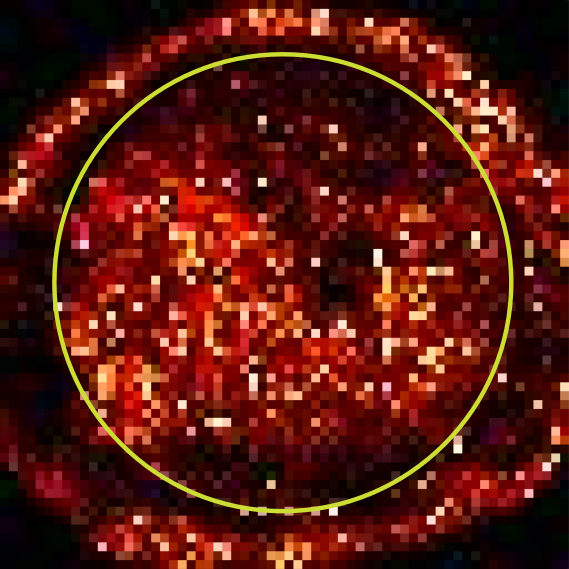}
    }\hfill
    \subfloat[$d = 7$]{%
        \includegraphics[width=0.23\linewidth]{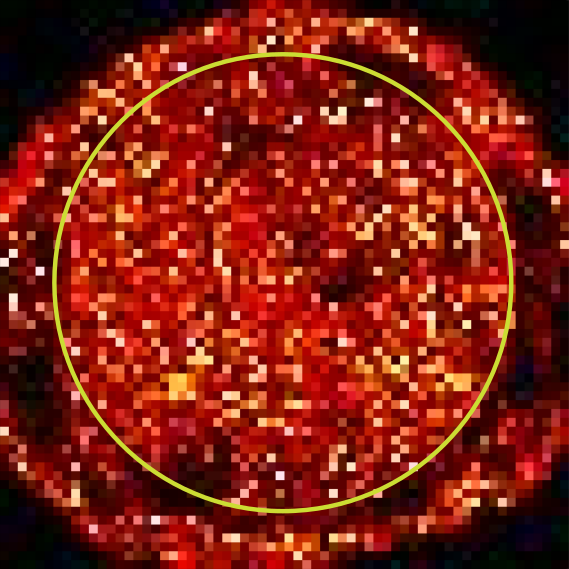}
    }\hfill
    \subfloat[$d = 20$]{%
        \includegraphics[width=0.23\linewidth]{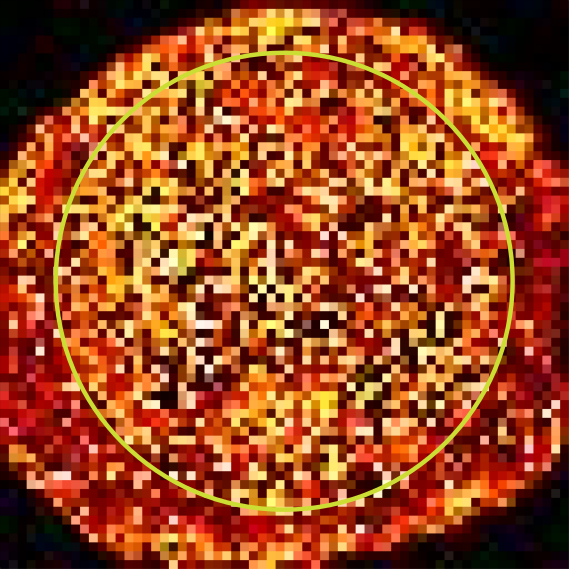}
    }\hfill
    \subfloat{%
        \includegraphics[width=\linewidth]{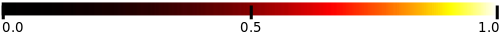}
    }
    \caption{An AIA image in JP2 format from $171$--{\AA} channel, and the heat-maps of Tamura directionality with different values for the variable $d$, where $t = 90\%$.}
    \label{fig:tdHeatmaps}
\end{figure}

To determine the domain for $d$, the minimum distance between the peaks, we should take a look at the histogram of angles. With $n$ bins, such a histogram can be generated as follows:
\begin{equation}
    h_{D} = \Bigl\{ \frac{N_{\theta}(k)}{\sum_{i=0}^{n-1}{N_{\theta}(i)}};\quad 0 \leq k \leq 2n-1\Bigr\}
\end{equation}
where $N_{\theta}(x)$ is the frequency of the angles within the interval $\Bigl[k \frac{\pi}{2n}, (k+1) \frac{\pi}{2n} \Bigr)$. Since what Tamura directionality targets is not the angle but the direction of the lines, the resultant histogram will be symmetric around $\theta = 0^{\circ}$. To avoid redundant computation, we consider only the angles within the interval $[0, 180^{\circ})$. Setting $n$ to $90$ gives us a histogram with the breaks at $0^{\circ}, 2^{\circ}, 4^{\circ}, \cdots,$ and $180^{\circ}$. For this domain of angles, the set $\mathcal{D} = \{1,3,5, \cdots, 29\}$ is an admissible domain for the minimum distance between two peak. Note that those values indicate the minimum distance (in number of bins) for a peak to have from an already identified peak, to be considered a dominant peak. In Fig.~\ref{fig:tdHeatmaps}, the heat-maps of Tamura directionality for three different settings of $d$ are shown.\par

\subsection{Building the Feature Space}\label{subsec:building}

For each of the four image parameters, we compute its feature space by calculating all different variations of that parameter on one month worth of $4k$ AIA images (January, $2012$). This is done on JP2, FITS, and clipped FITS images, separately.\par

\subsection{Dimensionality Reduction}\label{subsec:reduction}

To reduce the dimensionality of the computed feature spaces, the F-test in one-way analysis of variance (ANOVA) is used to pick the feature (per wavelength) which has the highest rank in separation of the three solar event-types, active region, coronal holes, and quiet sun. The score of each feature is computed as the ratio of between-group variability and within-group variability, where all the instances of each solar event type form a single group. The ranking procedure is as follows: for each feature, or setting, all the instances of the three event-types reported by HEK will be collected. Using random undersampling, we make sure that the number of instances in all three categories is the same to remedy the class-imbalance problem. After computing the features of interest on the image cells spanning the bounding boxes of events, the results will be summarized using the seven-number summary. With a ten-fold sampling, we use the F-test to rank the settings. We then aggregate the scores per setting on its seven-number summary, and finally sort the settings by their scores and report the highest per wavelength. As an example, the parameter Tamura directionality on JP2 AIA images in $94$--{\AA} wavelength channel, with $t = 25$ and $d = 1$, was ranked the best compared to any other variation of that image parameter. Table~\ref{table:bestSettings} summarizes the best setting per wavelength channel, for each of the three image formats.\par

\begin{figure}[t]
    \centering
    \includegraphics[width=1.0\columnwidth]{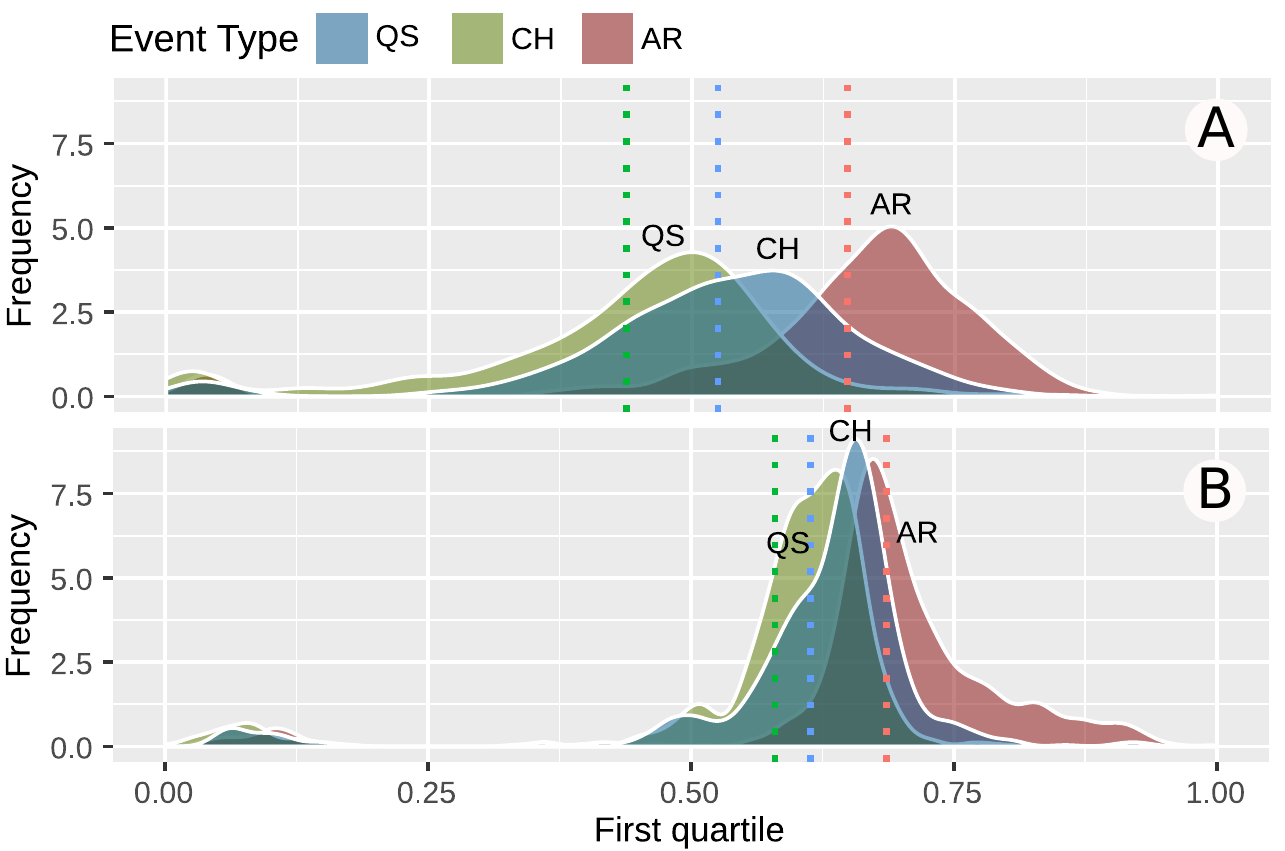}
    \caption{This plot illustrates the difference between the distribution of statistics of the best setting for an image parameter (A) and an arbitrary setting (B), on one month worth of $4K$ AIA images. The three colors distinguish the distributions of different solar event types (active region, coronal hole, and quit sun), and the dotted lines indicate the mean values of the distributions}. Note how in A the three distributions are more distinguishable. In this example, the image parameter is Tamura directionality, the wavelength is $94$--{\AA}, and the statistics is the first quartile.
    \label{fig:fscoreexample}
\end{figure}

To help understand how the best setting for an image parameter provides a better distinction between the instances of different event-types, an example is illustrated in Fig.~\ref{fig:fscoreexample}. In this visualization, the image parameter is Tamura directionality, and the chosen statistics is $Q1$ (first quartile). The difference between the distribution of $Q1$ of this parameter with the best setting as opposed to an arbitrary setting, on the three event types is shown. Note how in plot A, where the best setting is used, the three distributions are much more distinguishable compared to B where an arbitrary setting is used.\par

After this step, for each of the four image parameters, the dimensionality of the defined space shrinks down significantly, from several thousands to $63$ (for $9$ wavelength channels and $7$ summary statistics).\par

\begin{figure*}[!htp]
    \centering
    \includegraphics[width=1.0\linewidth]{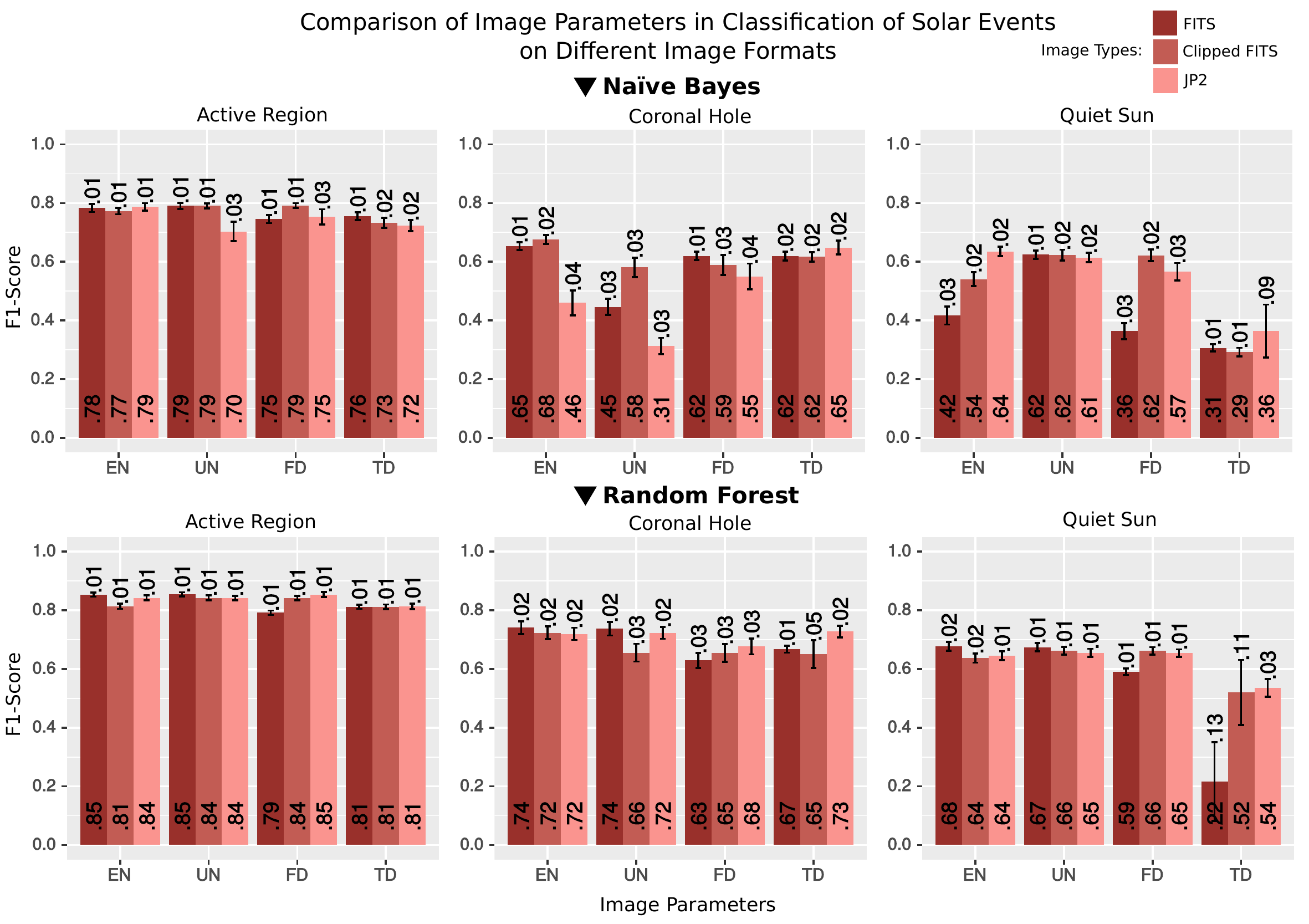}
    \caption{The classification results on the three event types (active region, coronal hole, and quiet sun) using Na{\"i}ve Bayes (first row) and Random Forest (second row) classifiers are illustrated here, separately for each event type using the f1-score measure. Each reported measure is averaged over $495$ trials of a $10$-fold cross validation sampling. Each trial was executed on a random sample of events' instances from  $13,518$ AR, $10,780$ CH, and $13,518$ QS event instances, within the period of 01-01-2012 through 31-12-2012. For each bar, the number on the bottom represents the f1-score value and the error interval shows the standard deviation of the f1-score. The image parameters are entropy (EN), uniformity (UN), Fractal Dimension (FD), and Tamura Directionality (TD).}
    \label{fig:f1scores}
\end{figure*}

\subsection{Building the Reduced Feature Space}\label{subsec:building2}

After reducing the dimensionality, the best setting for each image parameter is used to form the reduced feature space. This new feature space will then be generated based on one year (Jan 1 through Dec 31, 2012) worth of AIA images, for JP2, L1.5 FITS, and Clipped FITS images, with the cadence of $6$ minutes.\par

\subsection{Classification}\label{subsec:classification}

To measure the performance of the four image parameters after finding the best setting for each of them, we employ two classifiers, namely Na{\"i}ve Bayes and Random Forest\footnote{We use the Statistical Machine Intelligence and Learning Engine (smile) Java library: \url{http://haifengl.github.io/smile/}.}. Na{\"i}ve Bayes classifier \citep{maron1961automatic} is a simple statistical model that learns by applying the Bayes' theorem with strong independence assumption, on the labeled data and classifies based on the maximum a posteriori rule. In the context of our data points, for an event instance $e_t$ reported at time $t$, which can be of type $AR$, $CH$, or $QS$, it calculates the feature vector $v_t = \{x_1, \cdots, x_n\}$, where $n$ is the dimension of the defined feature space, and then classifies $e_t$'s event type, denoted by $\hat{y}_t$, as follows :
\begin{equation}
    \hat{y}_t = \argmax_{C_k \in \{AR, CH, QS\}}{p(C_k)\prod_{i=1}^{n}{p(x_i | C_k)}}    
\end{equation}

Since Na{\"i}ve Bayes classifier relies only on the probability of the occurrences of the events, the model is expected to perform poorly in classification of the less trivial cases. For the sake of completeness, we also employ Random Forest classifier \citep{ho1995random} for evaluation of the image parameters. This is an ensemble learning model that builds the decision trees on samples of data (a process called bootstrap aggregating) and classifies the class label by taking the majority vote of the trees classifying each data point. For our data, we generate a forest of $60$ different trees, each of which classifying the event types of the instances and at the end, the ensemble model makes the final decision by taking the majority vote of the trees.\par

For both classification models, we perform a $k$-fold cross-validation by sampling the events' instances on all combinations of any group of $4$ months in the year $2012$, resulting in $\binom{12}{4} = 495$ different trials. This allows having the test sets unbiased to the potential patterns in occurrence of solar events. Using repetitive random undersampling, we avoid the negative effect of imbalanced datasets as well.\par

For reporting the performance of these models we choose f1-score measure (also known as F-Score or F-Measure) which is the harmonic average of the precision and recall. Given precision $p$ be the number of correct positive classification divided by the total number of (correct or incorrect) positive results returned by the model, and recall $r$ be the number of correct positive classifications divided by the total number of instances of positive class, f1-score can be formulated as follows:
\begin{equation}
    \text{f1-score} = 2 \cdot (\frac{p \times r}{p + r}).
\end{equation}
Since we have three classes (AR, CH, and QS) for our classification models, f1-score should be reported for each class separately. To measure $p$ and $r$ for our ternary model, we use the one-against-all strategy which aims to classify an object of one type compared to the other two, whereas the one-against-one strategy would consider all pairs of classes and report the classification performance separately, which is unnecessary for our task. Furthermore, it is important to note that the undersampling step employed in the $k$-fold cross-validation provides balanced data for the models. Therefore, our choice of the performance measure does not need to be class-imbalance resistant, e.g., True Skill Score.\par

\begin{figure*}[htp] 
    \centering
    \includegraphics[width=1.0\linewidth]{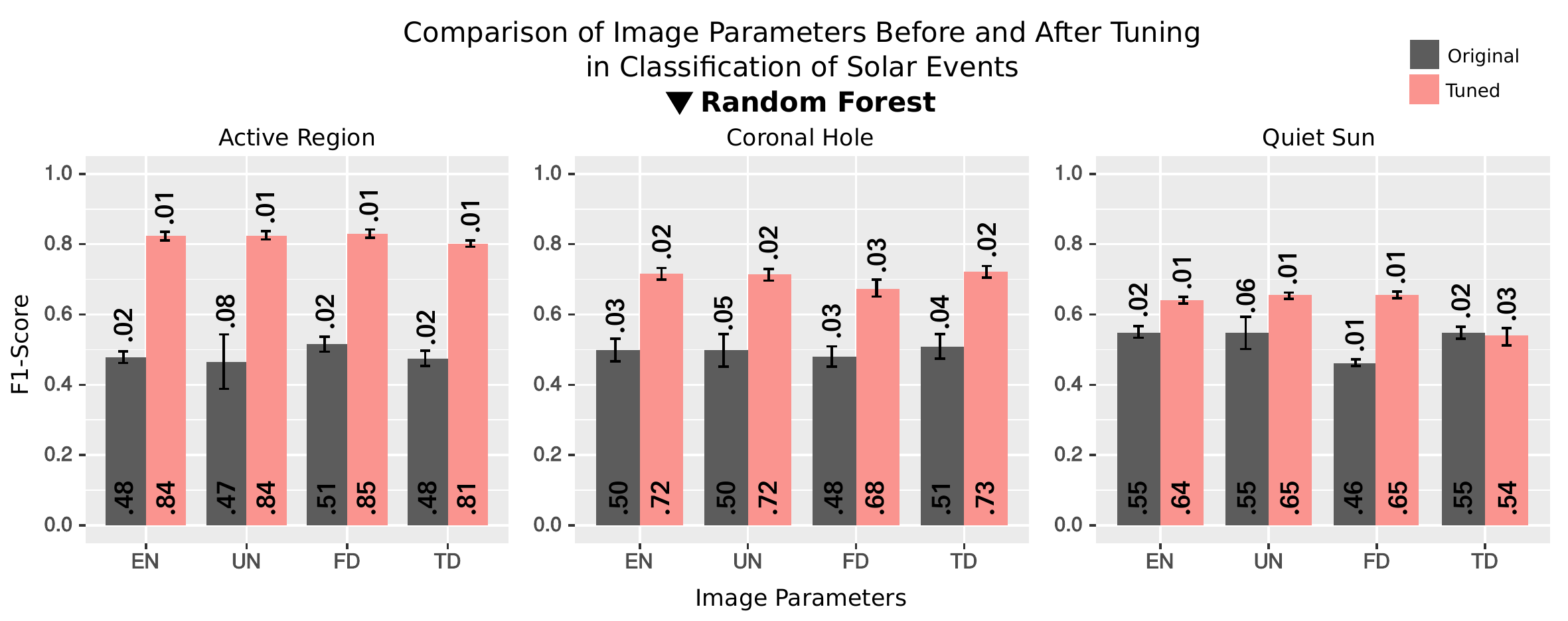}
    \caption{The illustration compares performances of Random Forest classifier in classification of three solar event types using each of the four image parameters, before and after tuning. The image parameters are entropy (EN), uniformity (UN), Fractal Dimension (FD), and Tamura Directionality (TD).}
    \label{fig:beforeAfterComparison}
\end{figure*}

The results of our experiments, using both Na{\"i}ve Bayes and Random Forest models, are illustrated in Fig.~\ref{fig:f1scores}. The key points about the results are enumerated below:
\begin{itemize}

    \item The performance of the two models is based on single image parameters and not their combinations. Random Forest, as we predicted before, performs significantly better. Using this model, one can observe that each of the four image parameters can individually classify active region instances fairly well ($\text{f1-score} > 0.8$) regardless of the image format. For the coronal hole instances, the results are only slightly lower but consistent ($\approx 0.7$ when JP2 images are used). The fact that such high confidence levels are reached using a set of very basic image parameters that are not domain specific (i.e., not tailored for classification of phenomena such as solar events) should stress the importance of our choices.
    
    \item Note that the relatively poor performance of both of the models in classification of QS is not a large concern, since it is just a synthesized event and some other event types that are reported to HEK but not used in this study could be adding noise to the instances labeled as QS, resulting lower purity in the class labels. However, the results are still above those expected if the samples were simply assigned a random label and therefore indicate the possibility that these parameters can transfer to other event type classification.  
    
    \item Another very important aspect of the results is in the comparison of the classification on different image formats, as the plots depict. For Random Forest classifier, in almost all cases, JP2 format is shown to be the better input for the model, compared to both FITS and clipped FITS. Even for Na{\"i}ve Bayes classifier which did not perform as well as Random Forest did, there is no consistent superiority when FITS or clipped FITS images were used compared to the JP2 format. This is despite the fact that FITS format theoretically contains more information than the compressed JP2, and therefore produces much larger files. In fact, an image in FITS format is $5$ to $14$ times larger than its JP2 version, depending on the wavelength channel used. With such understanding, we can now make our entire image repository $\approx10$ times smaller in size, with even some improvement in classification of solar events.
    
\end{itemize}

As one of our main contributions was to provide a dataset of tuned image parameters, we compare the classification of the solar events before and after the tuning steps on the image parameters. As shown in Fig.~\ref{fig:beforeAfterComparison}, our tuning results in significant improvement for all of the four image parameters across the event types. Note that the performance on the image parameters without tuning is only slightly above the random guess which is $0.33$. This is simply because the previous computation of the image parameters lack the thorough analysis of the individual parameters, and the tailored tuning steps.\par

Of course, the scope of this study is limited to tuning the image parameters, and the results in Fig.~\ref{fig:f1scores} and \ref{fig:beforeAfterComparison} reflect only the impact of the obtained image parameters, while better models (with higher performance or more robustness) can potentially be trained by exploring different classifiers, such as SVM or even deep neural networks, and tuning their hyper-parameters in a data-driven fashion.\par

\section{The Resultant Dataset}\label{sec:thedata}

Having demonstrated the effectiveness of utilizing tuned parameter settings for JP2 format AIA images, we then set out to produce a dataset ($\approx1$TiB/year) that is easily accessible for researchers wishing to utilize this data. The dataset we have created contains the ten image parameters listed in Table~\ref{table:tenParams}, which are processed from images captured by the SDO spacecraft, and are extracted from the AIA images at a six-minute cadence for each wavelength we process. As previously mentioned, the original images are high resolution ($4096\times4096$ pixel), full-disk snapshots of the Sun, taken in ten extreme ultraviolet channels (the nine channels that we utilize in this work are $94$\AA, $131$\AA, $171$\AA, $193$\AA, $211$\AA, $304$\AA, $335$\AA, $1600$\AA, and $1700$\AA) \cite{2012SoPh..275...17L}. The original high resolution images are accessible upon request from the Joint Science Operations Center, but our dataset is processed from the the JP2 compressed images available through the random access API at the Helioviewer repository\footnote{\url{https://api.helioviewer.org}}.\par

We have created an API\footnote{\url{http://dmlab.cs.gsu.edu/dmlabapi/}} that allows for the random access of the produced image parameter data. The processed dataset starts with observations from January 1, 2011 00:00:00 UTC and our intent is to continue to keep the dataset updated with the current observations for as long as the source of our data continues to provide new observations. The methods used for calculating the parameter values are released as part of our Open Source library DMLabLib\footnote{\url{https://bitbucket.org/gsudmlab/dmlablib}}. The settings for each of the parameter calculation methods that require some sort of setting value are listed in Table~\ref{table:bestSettings} of \ref{app:one}. Note that each of the nine waveband channels that we process has its own set of settings for each of the parameter calculation methods.\par

One already established use case for this dataset is tracking solar events that have been reported to the HEK \citep{kempton:phenomena, kempton2015tracking} where the parameters are used to perform visual comparisons of detections forming different possible paths a tracked event could take.  Another is the use of the parameters to perform whole image comparisons for similarity search in the context of content based image retrieval \citep{kempton:describing}. Similarly, the parameters have also been used to perform region comparison for similarity search in the context of region based content based image retrieval \citep{schuh2017region}. These are just a few of the possible use cases that we know have utilized a smaller and un-optimized previous version of this dataset. \ref{app:one} provides some additional analysis of the dataset produced by this work.\par

\section{Conclusion and Future Work}\label{sec:conclusion}
We presented the background information about the AIA images produced by the SDO mission and compared the FITS and JP2 image formats and the distribution of the pixel intensities in each of them. We also reviewed different aspects of each of the ten image parameters that we have selected to extract the important features of those images and then explained how we designed several different experiments to find the best settings for each of the features on different wavelength channels and the different image formats. After we obtained the best settings for each of the image parameters, we processed one year worth of data and extracted those features from the images queried with the cadence of $4$ hours. Finally, we presented our public dataset as an API by running several statistical analysis to illustrate a more accurate picture of the ready-to-use dataset.\par

We hope that our public dataset interests more researchers of different backgrounds and attracts more interdisciplinary studies to solar images. While we aim to keep our API data up-to-date with the stream of data coming from the SDO, we would like to expand it by adding more interesting image parameters, specifically computed for different solar events, which could lead to a better understanding of solar phenomena and higher classification accuracy.\par


\acknowledgments
This work was supported in part by two NASA Grant Awards [No. NNX11AM13A, and No. NNX15AF39G], and one NSF Grant Award [No. AC1443061]. The NSF Grant Award has been supported by funding from the Division of Advanced Cyberinfrastructure within the Directorate for Computer and Information Science and Engineering, the Division of Astronomical Sciences within the Directorate for Mathematical and Physical Sciences, and the Division of Atmospheric and Geospace Sciences within the Directorate for Geosciences. Also, we would like to mention that all images used in this work are courtesy of NASA/SDO and the AIA, EVE, and HMI science teams.\par




\appendix

\section{Statistical Analysis of Dataset}\label{app:one}

\begin{figure}[p]
    \vspace{1in}
    \includegraphics[width=\textwidth]{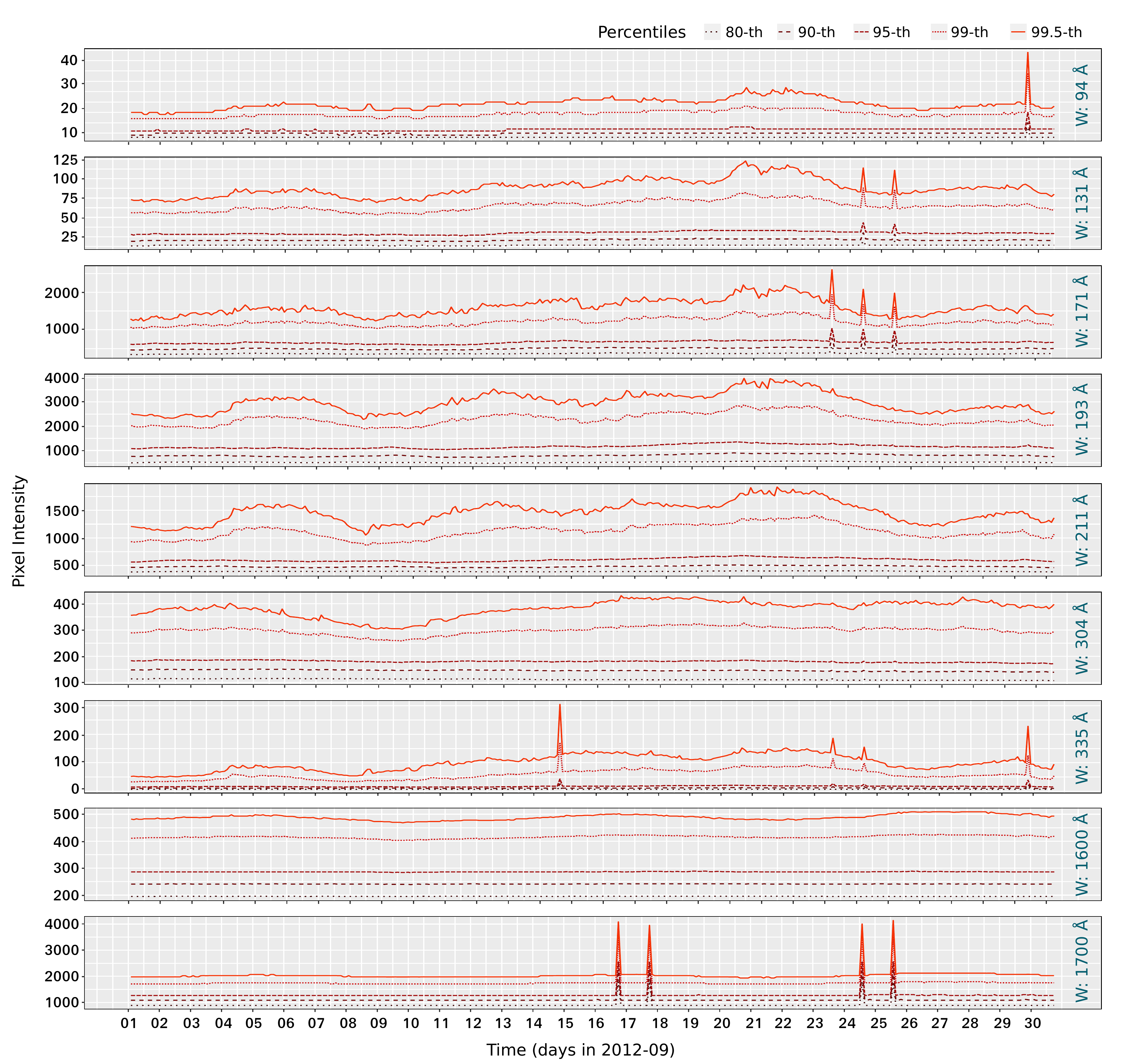}
    \caption{Different percentiles of pixel intensities for  $\approx 3240$ AIA FITS images (i.e., approximately $360$ images per wavelength channel). Each of the nine plots corresponds to one wavelength channel of the AIA instrument, specified in cyan, on the left. Each curve tracks the changes of the pixel intensity distribution of images captured every 2 hours, within the period of December $2012$.}
    \label{fig:percentiles}
\end{figure}

In this section, we present more statistical insight about the prepared dataset through a number of figures. Fig.~\ref{fig:percentiles} illustrates the changes in the distribution of pixel intensities of FITS images for the month of September 2012, with the cadence of 2 hours. We use this to support our argument for the cut-off point used in clipping of the FITS files in every wavelength channel (see Section~\ref{subsubsec:distribution}). Observing the changes of the $99.5$-th percentile of the pixel intensities in FITS images, knowing that several pixels with the maximum intensity value (i.e., $16383$) are present within this period, tells us that clipping at the highest point reached by this percentile while reducing the range of the intensities significantly, only affects $0.5\%$ of the pixels.\par

As an example, for images in $94$--{\AA} (see the first plot at the top of this figure), the highest value reached by the $99.5$-th percentile of the pixel values is equal to $44$ while pixels as bright as $16383$ are present. Among the five different percentiles, the one with the minimum effect on the images, i.e., $99.5$-th, is chosen for clipping of the FITS images to generate the new set of images that we referred to as \textit{clipped FITS}. The few sudden changes of the pixel intensities in Fig.~\ref{fig:percentiles} as we investigated, are mainly due to the several C-- and M-- class flares reported in this period. In some cases, the magnetically charged particles reaching the CCD detectors of the AIA instrument, also result in overexposed images, hence the spikes.\par

\begin{figure}[p]
    \vspace{1in}
    \includegraphics[width=\textwidth]{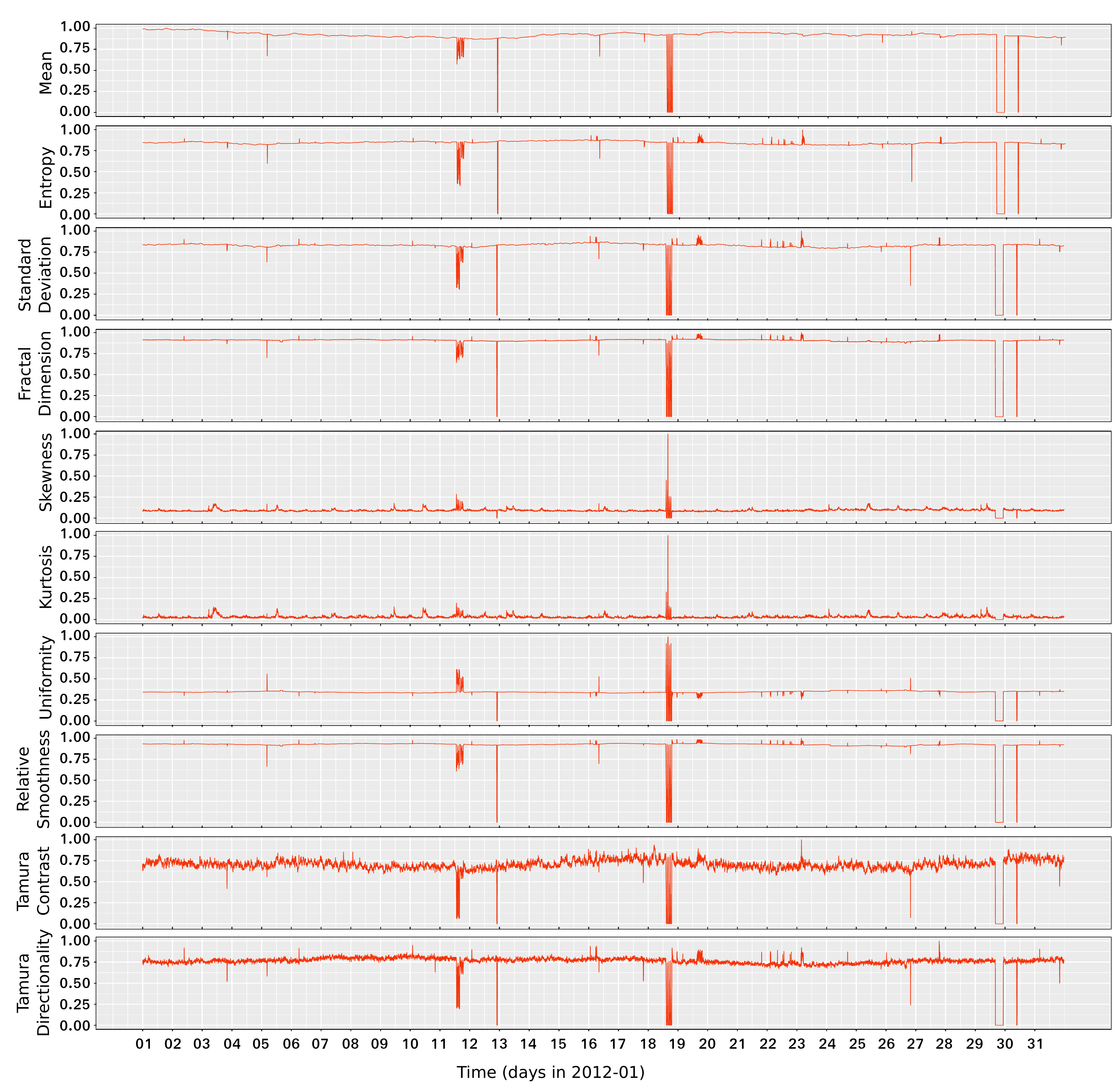}
    \caption{Mean of the ten image parameters extracted from images queried for a period of one month ($2012-01$). With the cadence of $6$ minutes, the plot represents $7440$ AIA images from the wavelength channel $171$--{\AA}.}
    \label{fig:meanofAll_2012}
\end{figure}

\begin{figure}[p]
    \vspace{.5in}
    \includegraphics[width=\textwidth]{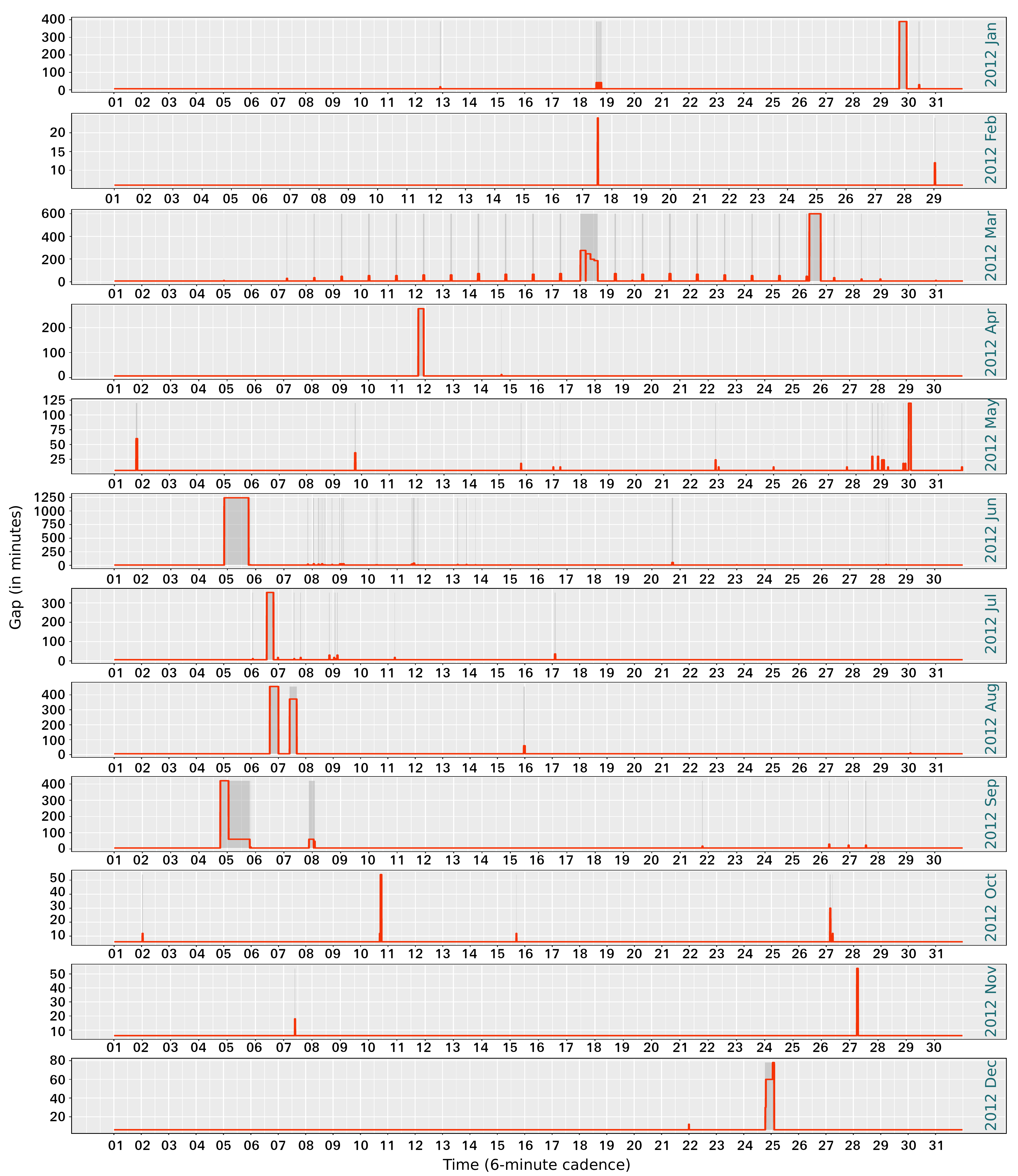}
    \caption{The time differences (in minutes) between image parameter files for AIA images, from the wavelength channel $171$--{\AA}, over the entire period of the year $2012$.}
    \label{fig:timeDiff_2012}
\end{figure}

\begin{figure}[p]
    \vspace{1in}
    \includegraphics[width=\textwidth]{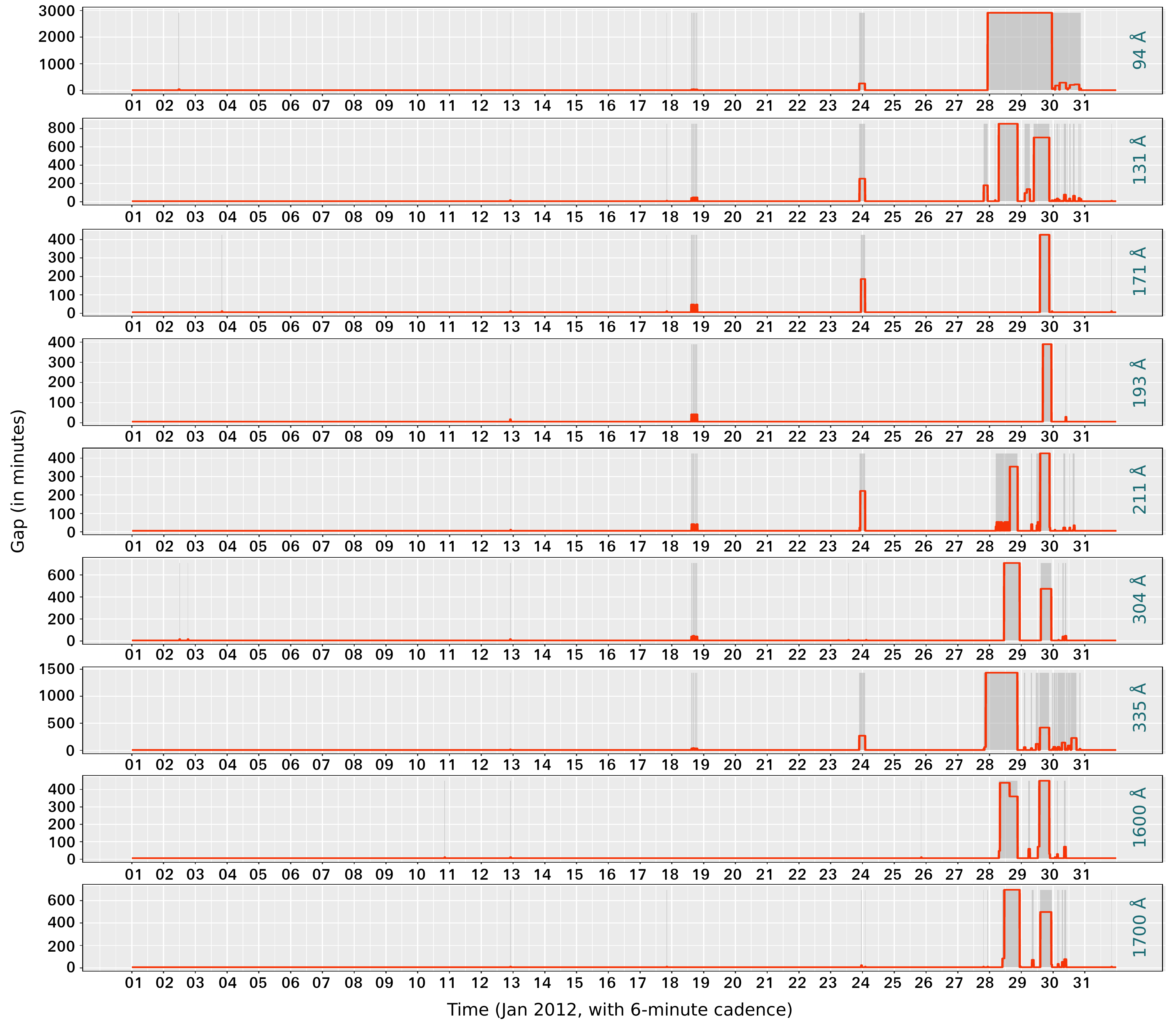}
    \caption{The time differences (in minutes) between image parameter files for AIA images, from the $9$ different wavelength channels, over the month of January $2012$.}
    \label{fig:timeDiff_jan}
\end{figure}

To present a big picture of the flow of data in the dataset, we show the mean value of each of the ten image parameters after they are extracted from the AIA images, for the entire month of January $2012$ (Fig.~\ref{fig:meanofAll_2012}). The ten image parameters for this plot are computed on the entire full-disk images and the mean statistics is then extracted from the resultant matrix. To present the continuity of the collected and computed data, we present the time differences between the image data points of our dataset, for the entire calendar year of $2012$, with the cadence of $6$ minutes, in Fig.~\ref{fig:timeDiff_2012} and for one month, across nine wavelength bands in Fig.~\ref{fig:timeDiff_jan}.\par

The small periods where the values go to zero in Fig.~\ref{fig:meanofAll_2012} are artifacts of missing input data and/or corrupted images that are uniformly black. Similarly, the periods where the time between reports peaks for some period is another indication of missing input data. This can be caused by any of numerous possible reasons that could cause a step in the processing pipeline to fail to receive an image from the previous step in the pipeline. These can range from the satellite not transmitting the data in the first place, to an error at any one of the processing steps prior to our processing of the JP2 image from Helioviewer. The missing data can also be caused, as found in \cite{schuh2015solar}, by the moon or earth itself occluding the view of the sun from the satellite on almost a daily basis, as seen in March $2012$ in Fig.~\ref{fig:timeDiff_2012}. In all, this does not represent a significant portion of the dataset given that the data corresponding to a few months in $2012$ are missing the largest portion compared to other years.\par

At the end, the best settings derived and used to generate this dataset is presented in Table.~\ref{table:bestSettings}. The numeric values mentioned in this table are mostly useful for the purpose of reproducibility of the dataset, since this is possible for those who find the creation steps of the dataset interesting, thanks to our open source library, DMLabLib\footnote{\url{https://bitbucket.org/gsudmlab/dmlablib}}.\par

\begin{table*}[p]
\vspace{1in}
\centering
\caption{\footnotesize{The best settings per wavelength, for the four image parameters across three image formats are listed here. In this table, $n$ indicates the number of bins used to compute entropy or uniformity, $t$ are $d$ are the threshold and peak-to-peak distance, respectively, used to measure directionality, and finally the variable $\sigma$ stands for the Gaussian smoothing parameter required in computing fractal dimension. For more details about these variables, see Section~\ref{subsec:settingsSummary}.}}
\begin{tabular}{@{} r r r r r r r r @{}}
\toprule
& \multicolumn{2}{c |}{Wavelength} & \multicolumn{1}{c |}{Uniformity} & \multicolumn{1}{c |}{Fractal Dimension} & \multicolumn{2}{c |}{Tamura Directionality} & \multicolumn{1}{c}{Entropy} \\ \toprule
 & \multicolumn{2}{c|}{(\AA)} & \multicolumn{1}{c|}{n} & \multicolumn{1}{c|}{sigma} & \multicolumn{1}{c|}{t} & \multicolumn{1}{c|}{d} & \multicolumn{1}{c}{n} \\ \cmidrule(l){1-8} 
 \parbox[t]{2mm}{\multirow{9}{*}{\rotatebox[origin=c]{90}{JP2}}}
 && 94 & 12 & 2.0 & 25 & 1 & 12  \\
 && 131 & 36 & 1.0 & 25 & 1 & 60  \\
 && 171 & 60 & 4.5 & 75 & 1 & 12 \\
 && 193 & 97 & 1.0 & 25 & 1 & 24 \\
 && 211 & 84 & 1.5 & 25 & 1 & 12 \\
 && 304 & 36 & 3.5 & 75 & 1 & 12 \\
 && 335 & 97 & 2.0 & 25 & 1 & 12 \\
 && 1600 & 109 & 2.5 & 90 & 1 & 12 \\
 && 1700 & 48 & 4.0 & 90 & 3 & 12 \\ \cmidrule(l){1-8} 
 \parbox[t]{2mm}{\multirow{9}{*}{\rotatebox[origin=c]{90}{Clipped FITS}}}
 && 94  & 62  & 4.5 & 7 & 5 & 104  \\
 && 131 & 1230& 4.0 & 7 & 4 & 175 \\
 && 171 & 3717& 4.5 & 9 & 3 & 1239 \\
 && 193 & 1889& 5.0 & 6 & 2 & 1889 \\
 && 211 & 796 & 2.0 & 9 & 4 & 796  \\
 && 304 & 615 & 5.0 & 9 & 4 & 615  \\
 && 335  & 1888& 4.0 & 7 & 4 & 435 \\
 && 1600 & 5090& 4.5 & 7 & 4 & 2666\\
 && 1700 & 1970& 3.0 & 4 & 3 & 1970 \\ \cmidrule(l){1-8}
 \parbox[t]{2mm}{\multirow{9}{*}{\rotatebox[origin=c]{90}{L1.5 FITS}}}
 && 94 & 12 & 4.0 & 25 & 21 & 3900\\
 && 131& 36 & 5.0 & 90 & 1  & 780 \\
 && 171& 60 & 0.0 & 25 & 23 & 780 \\
 && 193& 97 & 1.0 & 75 & 1  & 780 \\
 && 211& 84 & 1.0 & 75 & 1  & 780 \\
 && 304& 36 & 5.0 & 75 & 1  & 780 \\
 && 335& 97 & 4.0 & 25 & 21 & 2340 \\
 && 1600& 109& 5.0 & 90 & 3  & 780 \\
 && 1700& 48 & 3.5 & 25 & 23 & 780  \\ \cmidrule(l){1-8}
\end{tabular}
\label{table:bestSettings}
\end{table*}

\section{Impact of Non-zero Quality Observations}
In this section, we address the specific concern regarding the impact of the AIA instrument degradation, as well as usage of the ``low quality'' images, on our dataset. By ``low quality'' we mean images whose QUALITY flag in their header is set to a non-zero value \citep{nightingale2011aia}. This value is an integer whose $32$-bit binary representation describes $32$ different issues, such as missing flat-field data, missing orbit data, and the like.\par

\subsection{Impact of CCD Degradation}
The CCDs (charge coupled device) of the AIA instrument, like any electronic devices, are subject to degradation. The impact of CCD degradation was known prior to the launch of SDO \citep{boerner2011initial}, and has been studied ever since (e.g., \cite{fontenla2016five}). The effect of instrument degradation is a secular decrease over time in the data counts of the FITS files, which results in a gradual decrease in the pixel intensities of the AIA images. This trend, although is very subtle and only visible when the average data counts of FITS files are monitored over the course of several years, can potentially impact many pixel-based analyses of solar events (to the best of our knowledge, no study has provided sufficient evidence for such impact, and the characteristics of the tasks impacted are not clearly known). To this end, a periodic re-calibration of the instrument was planned prior to the launch of SDO and has been and will continue to be carried out periodically to ensure the quality of the data. The details of such calibration process is described in \cite{boerner2011initial}. Our dataset is based on the level $1.5$ data utilized by Helioviewer, whose gains are adjusted to use the above mentioned calibration so that there is a consistent ``zero level'' in the images.\par

In case the above procedure does not fully resolve the degradation impact, we still believe that the effect should be negligible to our dataset. This is mainly because of the different nature of our data points and the applications this dataset is meant to be used for. Specifically, the data points in our dataset are extracted image parameters, and not the raw pixel values. Furthermore, in this study, we were able to show that the extreme high end of the range of values in the recorded L1.5 FITS images are actually detrimental to results in our analysis, and therefore we are clipping these values. The clipping was done either in our pre-processing phase when we used the FITS files, or by Helioviewer's JP2GEN project that provided the JP2 images for our analyses. So, the dynamic range compression in the images that is introduced by having to turn up the gain as the CCD deteriorates will most likely not have a noticeable impact, if at all, on our work.

Additionally, the extracted parameters used in this study are minimally affected by the long-term global changes in image intensities, especially when applied to the clipped images. As an example, consider the standard deviation parameter from our dataset. This is computed in local regions of a processed image and the subtle changes of the overall dynamic range of the brightness of source images, caused by a drifting ``zero level'', will have minimal effect on the results when applied to images that are pre-processed using a clipping method to reduce the dynamic range of the intensity values. Another example would be fractal dimension, which is computed on the detected edges. As discussed in subsection \ref{subsubsec:fractalDimension}, the edge detection is carried out based on the local gradients within images, and therefore, mild long-term changes such as the one imposed by CCD degradation, will not have a significant impact on the computed dimension, if at all. Among the ten image parameters, only mean parameter is susceptible to the degradation. The magnitude of the impact can be determined by the degree of degradation that could not be completely resolved in the AIA level 1.5 data products.\par

\subsection{Impact of Instrument Anomalies}
Based on our empirical study of hundreds of AIA images with non-zero QUALITY values (i.e., low quality images), these images fall into two main groups. One comprises the images which are visually no different than any zero QUALITY AIA images. In fact, in some cases the missing information does not affect the pixel values of the images at all. The other group, however, contains images in which the Sun's disk is rotated, shifted, or blocked due to eclipse, or because of some instrumental artifacts, large patches of black squares appear on the images. These are certainly not proper inputs for any analyses.\par

To the best of our knowledge, the frequency of the $32$ different quality flags has not been studied on AIA images yet. Our brief study on several (non-consecutive) months worth of AIA images, with the cadence of $36$ seconds, shows presence of $\approx4.2\%$ of non-zero QUALITY images (both group one and two). Of course to achieve a reliable statistics as the fraction of low quality images on the entire AIA data collection, a much larger sample should be processed. But unfortunately, lack of proper documentation on the FITS keywords and absence of a publicly available database of the header information, makes it difficult to obtain a more thorough analysis on this topic. Therefore, we will leave the computation of a more comprehensive statistic on the fraction of images with fundamental quality issues (i.e., the second group), to the original AIA image data providers. Since we computed the ten image parameters on all AIA images that fell into our sampling cadence, regardless of their quality flag, we added the QUALITY value of images to our database, and provided the user with the corresponding requests to retrieve the QUALITY values from the API, as well as some other basic spatial header information which are needed for labeling of the solar events. It is up to the interested researchers to decide whether they prefer to keep the low quality images for their study or not.\par

It is worth noting that, regarding the first group of images, lack of some pieces of information may disqualify such images for some specific scientific analyses, however, we believe that machine learning models built on the extracted image parameters (i.e., our dataset) would not be effected by such unnoticeable differences. Preprocessing the raw data and achieving a cleaned dataset are indeed critical steps in any data-related analyses. This is, in fact, the premise of the current study. Having that said, machine learning models are designed to have a degree of resistance against noise. As they learn the global patterns and structures of the data by fitting mathematical models against a very large number of data points, and very often in a high-dimensional vector space, having a few data points with some additional noise in just a few dimensions, would not impact the overall performance of the models. This is our reasoning for not excluding the low-quality images. But users of the dataset can decide on this based on their understanding of the impact of low-quality images on their desired models.\par

\section{Impact of heterogeneous Exposure Time}
AIA is equipped with an automatic exposure control (AEC) which adjusts the length of time the cameras' sensors are exposed to light. This adjustment takes into account the overall brightness of the Sun. During occurrence of some solar activities such as large flares, some regions on the Sun are significantly brighter. In such cases, a shorter exposure time could produce an image of a higher quality. The exposure time used for each image is recorded in their header information. We use this information to normalize the pixel intensities of each image before we compute the image parameters.\par

\end{document}